\documentclass[aps,prx,reprint,superscriptaddress]{revtex4-2}

\usepackage{amsmath,amssymb,mathtools,bm,mathrsfs}
\usepackage{tikz}
\usetikzlibrary{arrows.meta,positioning}
\usepackage{graphicx}
\usepackage{dcolumn}
\usepackage{booktabs}
\usepackage{xcolor}
\usepackage[colorlinks=true,allcolors=blue!55!black]{hyperref}
\usepackage{microtype}
\newcommand{\dd}{\mathrm{d}}
\newcommand{\ii}{\mathrm{i}}

\newcommand{\Walg}{W_{1+\infty}}
\newcommand{\cH}{\mathcal{H}}

\newcommand{\cE}{\mathcal{E}}
\newcommand{\cA}{\mathcal{A}}
\newcommand{\cC}{\mathcal{C}}
\newcommand{\cI}{\mathcal{I}}
\newcommand{\cQ}{\mathcal{Q}}
\newcommand{\eps}{\varepsilon}

\begin{document}

\title{An Effective String Theory Toolbox for Quantum Hall Interfaces I: \\ Worldsheet Kinematics and Constraint Structure}

\author{Ken K. W. Ma}
\affiliation{Independent researcher, Orlando, Florida 32837, USA}
\date{\today}

\begin{abstract}
A freely moving quantum Hall interface is fundamentally different from an ordinary edge fixed by an external confining potential. Since a normal displacement changes the areas occupied by the adjacent incompressible phases, the interface geometry and charge dynamics cannot be treated as independent degrees of freedom.  We formulate this problem for interfaces between Abelian quantum Hall phases using a spatially reparametrization-invariant worldsheet description, in which tangential motion is a relabeling of the interface while normal motion is physical.  Starting from the two-sided Chern--Simons response, we derive the relation between normal charge transport and interface motion.  We then introduce a relative-area construction, defined with respect to a material reference curve, that converts this velocity relation into an equal-time constraint linking the charged boundary sector to the interface shape.  Combined with the folded $K$-matrix current algebra, this identifies the universal Hall kinematics of the moving interface while leaving its geometric energy and neutral dynamics dependent on microscopic interface physics.  The resulting framework provides a systematic basis for effective theories of dynamical quantum Hall interfaces.
\end{abstract}

\maketitle

\section{Introduction}
\label{sec:introduction}

The low-energy boundary of a quantum Hall (QH) liquid is usually described by
a one-dimensional edge fixed by an external confining potential.  Bulk
topology fixes the universal structure of the edge theory, whereas
electrostatics and interactions determine its propagation velocity and
higher-gradient corrections~\cite{Wen1990,Wen1991,WenBook,Stone1991}.  This
separation is appropriate for a pinned edge because its geometry is externally
prescribed rather than treated as a dynamical degree of freedom.

A freely moving interface is different in a fundamental way.  Its embedding
is dynamical, and a normal displacement changes the areas occupied by the
adjacent incompressible phases.  Charge and shape are therefore
kinematically linked.  For a self-bound Laughlin droplet, combining the
first-order dynamics fixed by the Hall response with line tension produces a
chiral mode with cubic rather than linear dispersion
~\cite{Giovanazzi1994,LiMa2021}.  The embedded-curve formulation of
Ref.~\cite{TurkerYang2022} organized the interface energy in terms of arc
length and curvature and reproduced the leading quintic correction.  The
remaining question is how to formulate the equal-time charge--shape
structure of a moving interface beyond the linearized theory.

This problem cannot be resolved by simply treating the chiral scalar and the
embedding as independent fields.  The bulk scalar mode carries an
Eulerian time derivative, whereas its pullback to a moving interface carries
a material derivative.  Confusing the two introduces a dependence on how the
boundary field is extended away from the interface and can obscure the
distinction between tangential relabeling and physical normal motion.
Related nonlinear edge and droplet theories have been obtained from
phase-space, lowest-Landau-level, and hydrodynamic reductions
~\cite{Polychronakos2005,KarabaliNair2004,MonteiroNairGaneshan2024}, but they do
not directly provide the local material charge--shape condition considered
here.

In this work, we formulate Abelian QH interfaces using a spatially
reparametrization-invariant worldsheet description.  Starting from the
two-sided Chern--Simons response, we derive the current jump condition at a
moving interface.  After choosing a material reference curve, a
relative-area construction provides an equal-time primitive of this velocity
relation and identifies the charged material density with the swept
magnetic-area density.  Combined with the folded $K$-matrix current algebra,
this determines the local material charge--shape condition and, about a straight
interface with $\Delta\nu\neq0$, yields the universal shape bracket.  Neutral
modes remain controlled by the full folded interface data and the allowed
gapping interactions.

We also study an auxiliary velocity-constrained formulation using the
Dirac--Bergmann procedure.  It shows that tangential motion generates spatial
reparametrizations, whereas normal motion is physical and belongs to a
second-class charged--shape sector.  The relative-area construction is
independent of a normal extension of the Chern--Simons scalar mode, although its
local density depends on the chosen material reference convention.

Here, our analysis is restricted to the electromagnetic Chern--Simons response in a
flat sample.  Wen--Zee and gravitational response terms may generate
additional geometric boundary contributions and are not included here
~\cite{WenZee1992,Read2009,GromovJensenAbanov2016}.  Furthermore, we assume screened or short-range interactions, a finite bulk gap, and an interface width small
compared with the local radius of curvature.  Unscreened Coulomb interactions
instead produce nonlocal boundary energetics and nonanalytic dispersion
~\cite{Stone1991,Giovanazzi1994}.

The manuscript is organized as follows. Section~\ref{sec:geometry} introduces the moving-interface geometry. Section~\ref{sec:symplectic} derives the Hall transport condition, and Sec.~\ref{sec:abelian} develops the Abelian material-sector construction. Sections~\ref{sec:canonical} and~\ref{sec:gaugefix} analyze the canonical constraints and the long-wavelength reduction.  Section~\ref{sec:nonlinear}
organizes the nonlinear geometric Hamiltonian and its observable
consequences.

\section{Worldsheet geometry of a moving interface}
\label{sec:geometry}

We first fix the geometric and orientation conventions for a moving planar interface, using standard results from differential geometry~\cite{doCarmo} and worldsheet formalism~\cite{Polchinski}. Let
\begin{equation}
\bm X:
(\tau,\sigma)
\longmapsto
\bm X(\tau,\sigma)\in\mathbb{R}^2
\label{eq:spatial_embedding}
\end{equation}
be a regular spatial embedding, $\bm X'\neq0$, where a prime denotes $\partial_\sigma$.  For a closed interface, $\sigma$ is periodic.  The induced metric and invariant line element are
\begin{equation}
\gamma=\bm X'^2,
\qquad
\dd s=\sqrt{\gamma}\,\dd\sigma,
\qquad
\partial_s=\gamma^{-1/2}\partial_\sigma.
\label{eq:metric_ds}
\end{equation}
We use $\eps_{12}=+1$, and define the unit tangent and outward normal as
\begin{equation}
\bm t=\partial_s\bm X,
\qquad
n^i=-\eps^{ij}t_j.
\label{eq:tangent_normal}
\end{equation}
The orientation of $\sigma$ is chosen so that $\bm n$ points from the QH fluid toward the other phase. With this convention,
\begin{equation}
\sqrt{\gamma}\,v_n
=
\eps_{ij}X'^i\dot X^j,
\label{eq:normal_area_element}
\end{equation}
where a dot denotes $\partial_\tau$ at fixed $\sigma$. The signed extrinsic curvature of the spatial curve is
\begin{equation}
K
=
\bm n\cdot\partial_s\bm t
=
\eps_{ij}t^i\partial_st^j
=
\frac{\eps_{ij}X'^iX''^j}{\gamma^{3/2}}.
\label{eq:signed_curvature}
\end{equation}
Its sign depends on the chosen orientation, while even powers of $K$ do not. It determines the Frenet equations~\cite{doCarmo}
\begin{equation}
\partial_s\bm t=K\bm n,
\qquad
\partial_s\bm n=-K\bm t.
\label{eq:frenet}
\end{equation}

The exact tension energy of the interface is
\begin{equation}
E_{\rm tens}[\bm X]
=T_0L[\bm X],
\qquad
L[\bm X]=\oint\dd s.
\label{eq:exact_tension}
\end{equation}
Higher-gradient terms are local scalars formed from $K$ and its arc-length derivatives.

\subsection{Normal and tangential motion}

Using the spatial embedding above, the interface velocity decomposes uniquely as
\begin{equation}
\dot{\bm X}=v_t\bm t+v_n\bm n,
\qquad
v_t=\dot{\bm X}\cdot\bm t,
\qquad
v_n=\dot{\bm X}\cdot\bm n.
\label{eq:velocity_decomposition}
\end{equation}
Normal motion changes the image of the curve in the physical sample, whereas tangential motion changes only the labels assigned to the same curve. For an infinitesimal active relabeling
\begin{equation}
\delta_\xi\bm X=\xi(\tau,\sigma)\bm X',
\label{eq:reparam_X}
\end{equation}
$\bm X$ transforms as a worldsheet scalar.  Direct differentiation gives
\begin{align}
\delta_\xi v_n
&=\xi v_n',
\\
\delta_\xi v_t
&=\xi v_t'+\sqrt{\gamma}\,\dot\xi.
\label{eq:velocity_reparam}
\end{align}
Thus $v_n$ is a scalar, whereas $v_t$ contains the inhomogeneous shift associated with time-dependent relabeling.

For a deformation
\begin{equation}
\delta\bm X=\eta_t\bm t+\eta_n\bm n,
\label{eq:general_deformation}
\end{equation}
the line element and curvature vary as
\begin{align}
\delta(\dd s)
&=
(\partial_s\eta_t-K\eta_n)\dd s,
\\
\delta K
&=
\partial_s^2\eta_n+K^2\eta_n+\eta_t\partial_sK.
\label{eq:variation_ds_K}
\end{align}
The corresponding time-evolution identities are collected in
Appendix~\ref{app:moving_geometry}. A derivation, including the variation of the arc-length derivative, is given in Appendix~\ref{app:geometric_variations}. These variations determine the normal functional derivative of a geometric Hamiltonian.  For
\begin{equation}
E_{0+2}
=
T_0\oint\dd s
+
T_2\oint\dd s\,K^2,
\label{eq:geometry_E02}
\end{equation}
a closed-curve variation yields
\begin{equation}
\delta E_{0+2}
=
\oint\dd s\,
\left[
-T_0K
+
T_2\left(2\partial_s^2K+K^3\right)
\right]\eta_n.
\label{eq:geometry_force}
\end{equation}
The first term is the normal energy gradient associated with line tension, while the second is the planar analogue of the bending contribution~\cite{Helfrich1973,LangerSinger1984}.  In a dissipative interface this force would drive curvature relaxation. In a QH fluid, it is instead converted into time evolution by the antisymmetric Poisson operator of the charged interface sector, turning a normal energy gradient into chiral propagation along the interface.  This is the reason that the same static energy functional leads to dynamics very different from ordinary capillary or elastic strings.

\subsection{Area, charge, and the static-gauge benchmark}

The same orientation fixes the signed enclosed area as
\begin{equation}
\cA[\bm X]
=
-\frac{1}{2}
\oint
\eps_{ij}X^i\dd X^j.
\label{eq:area_definition}
\end{equation}
Then
\begin{equation}
\delta\cA
=
\oint\dd s\,\eta_n,
\qquad
\frac{\dd\cA}{\dd\tau}
=
\oint\dd s\,v_n.
\label{eq:area_variation}
\end{equation}

For an interface between incompressible fluids of densities $\rho_1$ and $\rho_2$,
\begin{equation}
\delta N
=
\Delta\rho\,\delta\cA,
\qquad
\Delta\rho=\rho_1-\rho_2.
\label{eq:charge_area}
\end{equation}
At fixed magnetic field,
\begin{equation}
\Delta\rho
=
\frac{\nu_1-\nu_2}{2\pi\ell_B^2}.
\label{eq:density_difference}
\end{equation}
For a closed isolated droplet in the material sector, with fixed particle
number and $\Delta\rho\neq0$, one therefore has
\begin{equation}
\oint\dd s\,v_n=0,
\label{eq:fixed_area}
\end{equation}
which removes the uniform area-changing mode but not local shape modes. Moreover, we suppose the interface is the level set $F(\tau,\bm x)=0$ with $F<0$ inside. Then
\begin{equation}
\bm n=\frac{\bm\nabla F}{|\bm\nabla F|},
\qquad
v_n=-\frac{\partial_\tau F}{|\bm\nabla F|}
\quad\text{on }F=0.
\label{eq:level_set_velocity}
\end{equation}

\begin{figure}[htb]
\centering
\includegraphics[width=0.96\linewidth]{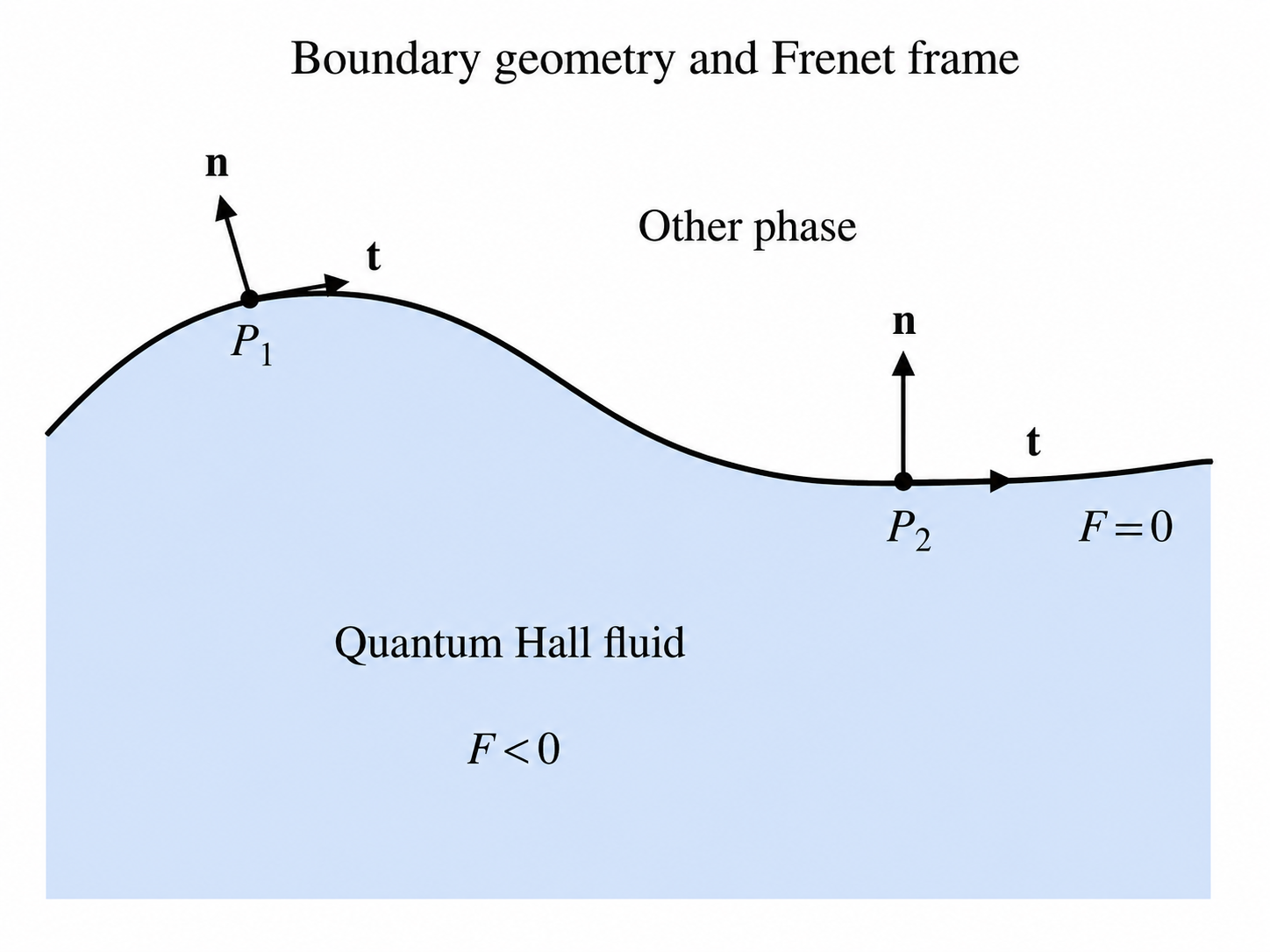}
\caption{Frenet frame and level-set description of a quantum Hall interface.
The solid curve is the level set $F=0$, the QH fluid occupies $F<0$, and
$\bm n$ points from the QH phase toward the other phase.}
\label{fig:Frenet_frame}
\end{figure}

The orientation convention, the Frenet frame, and the level-set description
are summarized in Fig.~\ref{fig:Frenet_frame}.

\subsection{Static-gauge benchmark}

For later comparison with the small-deformation theory, choose the local Monge gauge
\begin{equation}
\bm X(t,x)=(x,u(t,x)),
\label{eq:static_embedding}
\end{equation}
so that
\begin{align}
\dd s
&=
\sqrt{1+u_x^2}\,\dd x,
\\
\bm n
&=
\frac{(-u_x,1)}{\sqrt{1+u_x^2}},
\\
K
&=
\frac{u_{xx}}{(1+u_x^2)^{3/2}},
\\
v_n
&=
\frac{\dot u}{\sqrt{1+u_x^2}},
\\
v_t
&=
\frac{u_x\dot u}{\sqrt{1+u_x^2}}.
\label{eq:static_geometry}
\end{align}
The nonzero $v_t$ is fixed by the coordinate choice $X^1=x$ and is not a second physical mode. The small-slope expansions are then
\begin{align}
\dd s
&=
\left(1+\frac12u_x^2-\frac18u_x^4+\cdots\right)\dd x,
\\
K
&=
u_{xx}-\frac32u_x^2u_{xx}+\frac{15}{8}u_x^4u_{xx}+\cdots.
\label{eq:small_slope_geometry}
\end{align}
In particular,
\begin{equation}
E_{\rm tens}
=
T_0L_0
+\frac{T_0}{2}\int\dd x\,u_x^2
-\frac{T_0}{8}\int\dd x\,u_x^4
+\cdots,
\label{eq:tension_expansion}
\end{equation}
which reproduces the quadratic tension Hamiltonian of Ref.~\cite{LiMa2021}.

The geometric variables are now separated into a gauge direction, generated
by tangential relabeling, and a physical normal deformation.  The next step is
to determine how the Hall current couples to that normal motion.

\section{Hall transport on a moving domain}
\label{sec:symplectic}

We begin with a Laughlin fluid at $\nu=1/m$ occupying
\begin{align}
D(\tau)&=\{\bm x\mid F(\tau,\bm x)<0\},
\\
\Xi_D(\tau,\bm x)&=\Theta[-F(\tau,\bm x)].
\label{eq:moving_domain}
\end{align}
In units $e=\hbar=1$, the Abelian bulk action is~\cite{ZHK1989,FrohlichZee1991,WenBook,GromovJensenAbanov2016}
\begin{equation}
S_{\rm bulk}
=
\int\dd^3x\,\Xi_D
\left[
-\frac{m}{4\pi}
\eps^{\mu\nu\lambda}a_\mu\partial_\nu a_\lambda
+\frac{1}{2\pi}
\eps^{\mu\nu\lambda}A_\mu\partial_\nu a_\lambda
\right].
\label{eq:moving_CS}
\end{equation}
The physical current is
\begin{equation}
J^\mu
=
\frac{1}{2\pi}
\eps^{\mu\nu\lambda}\partial_\nu a_\lambda.
\label{eq:physical_current}
\end{equation}

Under the gauge transformation, $a_\mu\rightarrow a_\mu+\partial_\mu\alpha$, the electromagnetic coupling is unchanged and the Chern--Simons term varies by
\begin{equation}
\delta_\alpha S_{\rm bulk}
=
\frac{m}{4\pi}
\int\dd^3x\,
\alpha\,\partial_\mu\Xi_D\,
\eps^{\mu\nu\lambda}\partial_\nu a_\lambda,
\label{eq:gauge_variation_CS}
\end{equation}
up to temporal endpoint terms.  Since
\begin{equation}
\partial_\mu\Xi_D
=
-\delta(F)\partial_\mu F,
\label{eq:chi_derivative}
\end{equation}
the bulk gauge variation vanishes for unrestricted boundary values of $\alpha$ when
\begin{equation}
J^\mu\partial_\mu F=0
\qquad\text{on }F=0.
\label{eq:current_tangent}
\end{equation}
Using Eq.~\eqref{eq:level_set_velocity}, this becomes the exact local condition~\cite{TurkerYang2022}
\begin{equation}
J_n=\rho v_n.
\label{eq:moving_boundary_condition}
\end{equation}
For a uniform Laughlin fluid,
\begin{equation}
\rho=J^0=\frac{B}{2\pi m}.
\label{eq:Laughlin_density}
\end{equation}

Eq.~\eqref{eq:moving_boundary_condition} is simultaneously the kinematic condition for an incompressible moving boundary and the condition that cancels the gauge variation of the moving-domain bulk action.  This mechanism should be distinguished from the usual fixed-boundary anomaly-inflow construction.  For a pinned edge, the boundary cannot move to accommodate normal current and the charged boundary field remains an independent edge degree of freedom.  For an unpinned material interface, the same charged rearrangement can instead be represented by displacement of the boundary.  Thus, in the material sector the charged scalar and the shape describe the same low-energy charge transfer~\cite{Wen1990,GromovJensenAbanov2016,TurkerYang2022}. The action in Eq.~\eqref{eq:moving_CS} retains only the electromagnetic Chern--Simons response. Meanwhile, the Wen--Zee and gravitational sectors encode orbital spin, Hall viscosity, and thermal response, which can induce additional boundary-local terms involving intrinsic or extrinsic geometry~\cite{WenZee1992,Read2009,GromovJensenAbanov2016}.  Their inclusion is necessary for a complete geometric response theory, but it is logically separate from the charge-conservation condition studied here.

\subsection{Bulk scalar versus boundary pullback}

Variation with respect to $a_0$ gives the bulk Gauss law
\begin{equation}
m\eps^{ij}\partial_i a_j
=
\eps^{ij}\partial_i A_j=B.
\label{eq:Gauss_law}
\end{equation}
Locally in a simply connected droplet,
\begin{equation}
a_i=\frac{1}{m}A_i+\partial_i\Phi,
\label{eq:bulk_phi_solution}
\end{equation}
where $\Phi(t,\bm x)$ is a bulk gauge scalar.  Its pullback to the moving boundary is
\begin{equation}
\phi(t,\sigma)
=
\Phi[t,\bm X(t,\sigma)].
\label{eq:phi_pullback}
\end{equation}
The Eulerian derivative
\begin{equation}
\dot\Phi_E(t,\sigma)
\equiv
(\partial_t\Phi)[t,\bm X(t,\sigma)]
\label{eq:Eulerian_derivative}
\end{equation}
and the derivative of the pullback,
\begin{equation}
\dot\phi
=
\dot\Phi_E+\dot{\bm X}\cdot\bm\nabla\Phi,
\label{eq:material_derivative}
\end{equation}
are not the same.  The first is Eulerian, whereas the second differentiates the pullback at fixed boundary label.  This distinction is immaterial at leading order about a straight interface but is essential nonlinearly.

Adopting the boundary gauge $a_0|_\Sigma=0$ used in Ref.~\cite{TurkerYang2022}, and taking the background field to be time independent, the normal current can be written, with the orientation chosen above, as
\begin{equation}
J_n
=
\frac{1}{2\pi\sqrt{\gamma}}
\partial_\sigma\dot\Phi_E.
\label{eq:Jnormal_Eulerian}
\end{equation}
The moving-boundary condition therefore becomes
\begin{equation}
\partial_\sigma\dot\Phi_E
=
\frac{B}{m}\sqrt{\gamma}\,v_n.
\label{eq:implicit_constraint}
\end{equation}
For $m=1$ and $B=1$, Eqs.~\eqref{eq:Jnormal_Eulerian} and \eqref{eq:implicit_constraint} reduce to Eqs.~(24) and (28) of Ref.~\cite{TurkerYang2022}. Accordingly, the reduced Chern--Simons term is
\begin{equation}
S_{\rm top}
=
-\frac{m}{4\pi}
\int\dd t\,\dd\sigma\,
\phi'\dot\Phi_E,
\label{eq:implicit_topological_action}
\end{equation}
up to background-only and total-derivative terms.  Equations~\eqref{eq:implicit_constraint} and \eqref{eq:implicit_topological_action} show why the exact topological action is an implicit functional of the embedding.

A useful but nonunique extension gauge sets the normal derivative of $\Phi$ to zero at the boundary.  In that gauge,
\begin{equation}
\dot\Phi_E
=
\dot\phi-\beta\phi',
\qquad
\beta=\frac{v_t}{\sqrt{\gamma}}
=
\frac{\dot{\bm X}\cdot\bm X'}{\gamma}.
\label{eq:normal_extension_gauge}
\end{equation}
The combination on the right transforms as a scalar under time-dependent relabelings. Eq.~\eqref{eq:normal_extension_gauge} is a choice of how the boundary field is extended off the interface, not an additional physical law.  Any nonlinear canonical construction based on it must demonstrate independence of that extension choice.

Eq.~\eqref{eq:implicit_constraint} is a relation among velocities.  It
cannot yet be imposed as an equal-time constraint on the phase space of
$\phi$ and $\bm X$.  The relative-area construction in the next section
provides the required primitive without choosing how $\Phi$ extends away from
the interface.

\section{Two-sided Abelian interfaces and the material-sector condition}
\label{sec:abelian}

We now pass from a Laughlin--vacuum boundary to an interface between two Abelian phases.  The two-sided theory fixes the charged current algebra, while the relative-area construction supplies an equal-time primitive in a fixed material convention.

Let phase $r=1,2$ be described by an integral, symmetric, nondegenerate $K$ matrix $K^{(r)}$ and charge vector $\bm t^{(r)}$~\cite{WenZeeKMatrix1992,WenBook}.  Its filling fraction is
\begin{equation}
\nu_r
=
\bm t^{(r)T}
\bigl(K^{(r)}\bigr)^{-1}
\bm t^{(r)}.
\label{eq:abelian_nu_r}
\end{equation}
Take phase 1 to occupy $F<0$ and phase 2 to occupy $F>0$, with indicators $\Xi_1=\Theta(-F)$ and $\Xi_2=\Theta(F)$.  The electromagnetic Abelian Chern--Simons action is
\begin{equation}
\begin{aligned}
S_{\rm bulk}^{(1|2)}
=
\sum_{r=1}^{2}
\int\dd^3x\,\Xi_r
\Bigl[&
-\frac{1}{4\pi}
K_{IJ}^{(r)}
\eps^{\mu\nu\lambda}
a_\mu^{(r)I}\partial_\nu a_\lambda^{(r)J}
\\
&+\frac{1}{2\pi}
t_I^{(r)}
\eps^{\mu\nu\lambda}
A_\mu\partial_\nu a_\lambda^{(r)I}
\Bigr].
\end{aligned}
\label{eq:abelian_two_side_CS}
\end{equation}
The physical current in phase $r$ is
\begin{equation}
J_r^\mu
=
\frac{1}{2\pi}
t_I^{(r)}
\eps^{\mu\nu\lambda}
\partial_\nu a_\lambda^{(r)I}.
\label{eq:abelian_phase_current}
\end{equation}
If no singular line current is stored independently at the interface, the distributional current is
\begin{equation}
J_{\rm tot}^\mu
=
\Theta(-F)J_1^\mu
+
\Theta(F)J_2^\mu.
\label{eq:abelian_total_current}
\end{equation}
Using $\partial_\mu J_r^\mu=0$ in the two bulks gives
\begin{equation}
\partial_\mu J_{\rm tot}^\mu
=
\delta(F)
\bigl(J_2^\mu-J_1^\mu\bigr)
\partial_\mu F.
\label{eq:abelian_distributional_divergence}
\end{equation}
Distributional charge conservation therefore gives
\begin{equation}
J_{1,n}-J_{2,n}
=
(\rho_1-\rho_2)v_n.
\label{eq:abelian_jump_condition}
\end{equation}
At fixed magnetic field,
\begin{equation}
\rho_r=\frac{B}{2\pi}\nu_r,
\qquad
\Delta\nu=\nu_1-\nu_2,
\qquad
\Delta\rho=\frac{B}{2\pi}\Delta\nu.
\label{eq:abelian_density_jump}
\end{equation}
If the interface carries independently stored line charge, define
$q_{\rm ex}(\tau,\sigma)\,\dd\sigma$ as the excess charge in a coordinate
interval and $j_{\rm ex}(\tau,\sigma)$ as its flux through a line of fixed
$\sigma$.  The jump condition is then replaced by the worldsheet continuity
equation
\begin{equation}
\partial_\tau q_{\rm ex}
+
\partial_\sigma j_{\rm ex}
=
\sqrt\gamma
\left[
J_{1,n}-J_{2,n}-\Delta\rho\,v_n
\right].
\label{eq:excess_line_continuity}
\end{equation}
Here $j_{\rm ex}$ is the excess tangential current associated with the
independently stored line charge.  The material-interface sector is
$q_{\rm ex}=j_{\rm ex}=0$.

\subsection{Local semiclassical reduction of the two-sided bulk theory}

For a prescribed smooth embedding, the local Chern--Simons reduction begins with
\begin{equation}
Z[\bm X,A]
=
\int
\prod_{r=1}^{2}\mathcal D a^{(r)}
\exp\left\{
\ii S_{\rm bulk}^{(1|2)}
\right\}.
\label{eq:two_side_path_integral}
\end{equation}
Integration over the temporal components imposes the Chern--Simons Gauss laws
\begin{equation}
K_{IJ}^{(r)}
\eps^{ij}\partial_i a_j^{(r)J}
=
t_I^{(r)}B.
\label{eq:abelian_Gauss_law}
\end{equation}
Locally in a simply connected patch,
\begin{equation}
a_i^{(r)I}
=
\bigl(K^{(r)-1}\bm t^{(r)}\bigr)^I A_i
+
\partial_i\Phi_r^I,
\label{eq:abelian_Gauss_solution}
\end{equation}
up to harmonic modes.  Substituting the local solution into the first-order bulk action isolates the boundary symplectic term~\cite{KarabaliNair2004,Polychronakos2005}
\begin{equation}
S_{\Sigma}^{(1)}
=
-\frac{1}{4\pi}
\int\dd\tau\,\dd\sigma\,
\bm\phi'^{T}K_\Sigma\dot{\bm\Phi}_{E}
+S_{\rm bg}+S_{\rm hol},
\label{eq:bulk_reduced_boundary_action}
\end{equation}
where $S_{\rm bg}$ contains background-field terms, $S_{\rm hol}$ contains compact holonomies and topological sectors. Also,
\begin{equation}
K_\Sigma
=
K^{(1)}\oplus[-K^{(2)}],
\qquad
\bm t_\Sigma
=
\bm t^{(1)}\oplus\bm t^{(2)}.
\label{eq:folded_K_matrix}
\end{equation}
These are the folded interface data, for which the original interface
is treated as a boundary of the product theory~\cite{KapustinSaulina2011, BarkeshliJianQi2013, LanWangWen2015}.

The vector $\bm\phi$ denotes the pullbacks of the two bulk gauge scalars, with the second side written in folded orientation.  Eq.~\eqref{eq:bulk_reduced_boundary_action} is a local semiclassical reduction of the pure Chern--Simons sector.  The functional determinant and global sums affect local counterterms, compact sectors, and global state counting. The classical local first-order term retains the level $K_\Sigma$.

In the fixed electromagnetic background convention used here, the electromagnetic line density is
\begin{equation}
q_\Sigma
=
\frac{1}{2\pi}
\bm t_\Sigma^T\bm\phi'.
\label{eq:abelian_line_density}
\end{equation}
The induced Dirac bracket is
\begin{equation}
\{q_\Sigma(x),q_\Sigma(y)\}_D
=
-\frac{\Delta\nu}{2\pi}
\partial_x\delta(x-y).
\label{eq:abelian_charge_KM}
\end{equation}
This is the charged $U(1)$ Kac--Moody algebra of the folded
$K$-matrix edge theory~\cite{Wen1990,Wen1991,WenBook}, where
\begin{equation}
\bm t_\Sigma^T K_\Sigma^{-1}\bm t_\Sigma
=
\Delta\nu.
\label{eq:folded_anomaly_coefficient}
\end{equation}
Thus the charged current algebra depends only on the Hall-conductance difference, even when the interface carries additional neutral fields.

\subsection{Relative-area primitive and the material-sector condition}

The velocity jump condition admits an equal-time primitive after a material
reference convention is chosen.  Let $\bm X_0(\sigma)$ be a fixed reference
loop with the same labeling as the physical interface, and let
\begin{equation}
\bm Y(\tau,\sigma,r),
\qquad
0\le r\le1,
\label{eq:relative_area_homotopy}
\end{equation}
be a smooth interpolation satisfying
$\bm Y(\tau,\sigma,0)=\bm X_0(\sigma)$ and
$\bm Y(\tau,\sigma,1)=\bm X(\tau,\sigma)$.  Define
\begin{equation}
\mathfrak a_a
=
-B\int_0^1\dd r\,
\eps_{ij}
\partial_rY^i\partial_aY^j,
\qquad
a=\tau,\sigma.
\label{eq:relative_area_connection}
\end{equation}
A direct differentiation gives
\begin{equation}
\partial_\tau\mathfrak a_\sigma
-
\partial_\sigma\mathfrak a_\tau
=
B\sqrt\gamma\,v_n,
\label{eq:relative_area_curvature}
\end{equation}
and
\begin{equation}
\oint\dd\sigma\,\mathfrak a_\sigma
=
B\left[
\cA[\bm X]-\cA[\bm X_0]
\right].
\label{eq:relative_area_integral}
\end{equation}
The derivation of Eqs.~\eqref{eq:relative_area_curvature}
and~\eqref{eq:relative_area_integral}, together with the
exact-shift property in Eq.~\eqref{eq:relative_area_exact_shift},
is given in Appendix~\ref{app:relative_area}.

For homotopies with the same endpoints in the contractible plane,
$\mathfrak a_a$ changes by an exact one-form,
\begin{equation}
\mathfrak a_a\longrightarrow
\mathfrak a_a+\partial_a\Lambda.
\label{eq:relative_area_exact_shift}
\end{equation}
Consequently, the integrated relative area is invariant, whereas the local
density $\mathfrak a_\sigma$ depends on the material reference pairing and
interpolation convention.  This is analogous to the dependence of a sharp
interface density on a chosen dividing surface.  We keep that convention
fixed in what follows.

\begin{figure*}[t]
\centering
\includegraphics[width=0.7\textwidth]{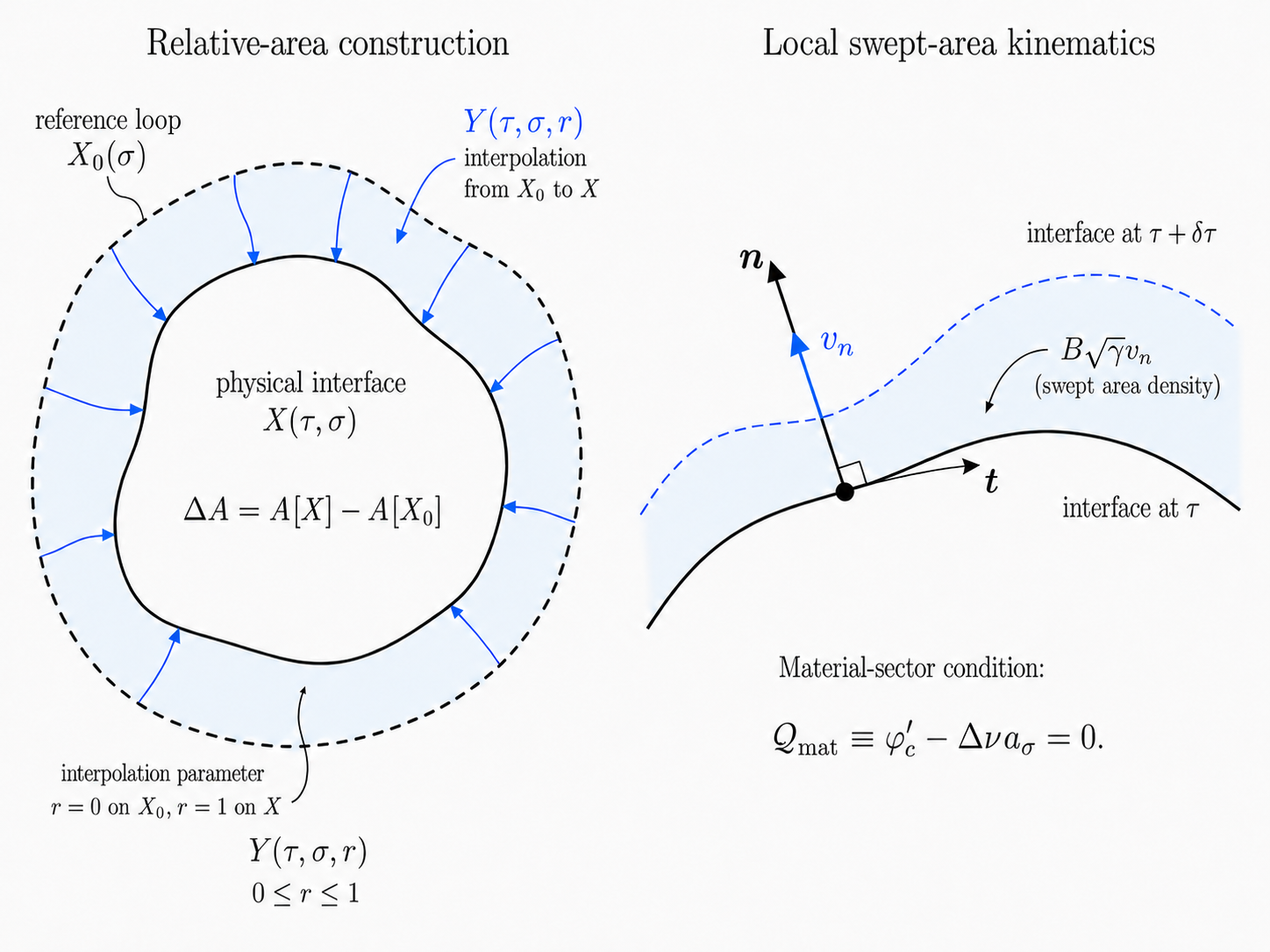}
\caption{Relative-area construction and local swept-area kinematics.
Left: a fixed material reference loop $\bm X_0(\sigma)$ is connected to the
physical interface $\bm X(\tau,\sigma)$ by a smooth interpolation
$\bm Y(\tau,\sigma,r)$, defining the relative magnetic-area density
$\mathfrak a_\sigma$.  Right: normal motion sweeps magnetic area at the local
rate $B\sqrt\gamma\,v_n$, leading in the material sector to
$\mathcal Q_{\rm mat}\equiv\varphi_c'-\Delta\nu\,\mathfrak a_\sigma = 0$.
The local density depends on the chosen material pairing and interpolation,
whereas its integral gives the signed relative magnetic area
$B[\mathcal A[\bm X]-\mathcal A[\bm X_0]]$.}
\label{fig:relative_area_construction}
\end{figure*}

Define the charged boundary field
\begin{equation}
\varphi_c
=
\bm t_\Sigma^T\bm\phi.
\label{eq:charged_field_definition}
\end{equation}
In the material sector, all interfacial charge is accounted for by displacement
of the two incompressible bulk densities.  In a fixed material convention the
equal-time condition is therefore
\begin{equation}
\cQ_{\rm mat}
\equiv
\varphi_c'
-
\Delta\nu\,\mathfrak a_\sigma
=0.
\label{eq:material_charge_shape_constraint}
\end{equation}
Its integral gives
\begin{equation}
\frac{1}{2\pi}
\oint\dd\sigma\,\varphi_c'
=
\Delta\rho
\left[
\cA[\bm X]-\cA[\bm X_0]
\right].
\label{eq:material_total_charge}
\end{equation}
Taking a time derivative and using
Eq.~\eqref{eq:relative_area_curvature} yields
\begin{equation}
\partial_\sigma
\left(
\dot\varphi_c-\Delta\nu\,\mathfrak a_\tau
\right)
=
B\Delta\nu\sqrt\gamma\,v_n.
\label{eq:intrinsic_moving_condition}
\end{equation}
Thus Eq.~\eqref{eq:material_charge_shape_constraint} is an equal-time primitive
of the two-sided velocity jump condition in the chosen material convention.
It does not require a normal extension of the bulk scalar.  The relative-area
construction and its local swept-area interpretation are summarized in
Fig.~\ref{fig:relative_area_construction}.

The convention-independent local quantity is the excess charge relative to
the displaced bulk charge,
\begin{equation}
q_{\rm ex}
=
\frac{1}{2\pi}
\left(
\varphi_c'-\Delta\nu\,\mathfrak a_\sigma
\right).
\label{eq:invariant_excess_density}
\end{equation}
A change of interpolation can be accompanied by the coordinate redefinition
$\varphi_c\to\varphi_c+\Delta\nu\Lambda$, under which
$q_{\rm ex}$ is unchanged.  The material sector is $q_{\rm ex}=0$.
The individual quantity $\varphi_c'/(2\pi)$ should therefore be interpreted as
the charged coordinate density in the chosen convention, not as a
reference-independent local observable.
These statements are local and classical within a fixed winding and total-charge
sector.  For a closed compact interface, the integrated condition must also be
compatible with the compactification and charge lattice of the folded theory.

We emphasize an important limitation.  The relative-area identity and the
equal-time condition do not by themselves determine a unique nonlinear
first-order action.  In particular, the naively covariantized chiral kinetic
term is not invariant off shell under
$\mathfrak a\to\mathfrak a+\dd\Lambda$ without additional background
boundary terms.  Those terms depend on the complete electromagnetic boundary
reduction and on the chosen microscopic dividing surface.  We therefore do
not use such an action in the remainder of the paper.  The finite-deformation
result retained here is the kinematic condition
Eq.~\eqref{eq:material_charge_shape_constraint}; the explicit reduced shape
bracket is derived only after linearization.

\subsection{Charged--neutral decomposition and interface Hamiltonian}

For $\Delta\nu\neq0$, define the charged direction
\begin{equation}
\bm w
=
\frac{K_\Sigma^{-1}\bm t_\Sigma}{\Delta\nu},
\qquad
\bm t_\Sigma^T\bm w=1.
\label{eq:charged_direction}
\end{equation}
Then decompose locally over the real fields
\begin{equation}
\bm\phi
=
\bm w\,\varphi_c
+
\bm\phi_n,
\qquad
\bm t_\Sigma^T\bm\phi_n=0.
\label{eq:charged_neutral_decomposition}
\end{equation}
Because $\bm w^TK_\Sigma\bm\phi_n=0$, the local folded symplectic term separates algebraically as
\begin{equation}
S_{\Sigma}^{(1)}
=
-\frac{1}{4\pi\Delta\nu}
\int\varphi_c'\dot\varphi_c
-\frac{1}{4\pi}
\int\bm\phi_n'^TK_n\dot{\bm\phi}_n,
\label{eq:charged_neutral_topological_action}
\end{equation}
for a prescribed boundary and in a fixed boundary convention, where $K_n$ is the restriction of $K_\Sigma$ to the neutral subspace.  This decomposition is local in field space, whereas compactification and the integral anyon lattice can obstruct a globally integral charged--neutral basis.

A leading local interface Hamiltonian is $H_\Sigma=\int\dd\sigma\,\cH_\Sigma$, with density
\begin{equation}
\cH_\Sigma
=
\sqrt\gamma
\left[
\frac{1}{4\pi}
V_{AB}\partial_s\phi^A\partial_s\phi^B
+U_{\rm gap}(\bm\phi)
+\cdots
\right],
\label{eq:general_interface_Hamiltonian}
\end{equation}
with a positive matrix $V$ in stable gapless sectors.  Possible gapping interactions are
\begin{equation}
U_{\rm gap}
=
-\sum_a g_a
\cos(\bm\ell_a^T\bm\phi),
\label{eq:abelian_gapping_potential}
\end{equation}
and a mutually commuting, charge-conserving set obeys
\begin{align}
\bm\ell_a^TK_\Sigma^{-1}\bm\ell_b
&=0,
\label{eq:abelian_null_condition}
\\
\bm t_\Sigma^TK_\Sigma^{-1}\bm\ell_a
&=0
\label{eq:abelian_charge_neutrality}
\end{align}
for all condensed vectors~\cite{Haldane1995,SantosHughes2017,MayMannHughes2019}.  In addition, each $\bm\ell_a$ must represent a microscopically allowed local tunneling operator in the folded anyon lattice.  To gap an entire nonchiral sector, the set must also contain a sufficient number of independent primitive vectors compatible with the signature and compactification lattice of $K_\Sigma$.  If all neutral modes are gapped or dynamically decoupled, the material constraint gives
\begin{equation}
q_\Sigma
=
\frac{1}{2\pi}\varphi_c'
=
\Delta\rho\,u
\label{eq:abelian_displacement_density}
\end{equation}
in linear static gauge, and hence
\begin{equation}
\{u(x),u(y)\}_D
=
-\frac{2\pi}{B^2\Delta\nu}
\partial_x\delta(x-y).
\label{eq:abelian_shape_bracket}
\end{equation}
The corresponding geometric dispersion is
\begin{equation}
\omega(k)
=
\frac{2\pi}{B^2\Delta\nu}
\left(
T_0k^3+2T_2k^5+\cdots
\right),
\label{eq:abelian_dispersion}
\end{equation}
up to orientation.  If protected neutral modes remain, they must be retained as additional physical worldsheet fields.  When $\Delta\nu=0$, the net Hall-induced charged block is degenerate, and any shape symplectic form must arise from neutral, geometric-response, or non-topological data.

Equations~\eqref{eq:material_charge_shape_constraint} and
\eqref{eq:charged_neutral_topological_action} summarize the local charged-sector kinematics in a fixed material convention while leaving neutral modes as separate physical fields.  To expose the canonical constraint algebra and to
retain independently stored line charge, we next introduce a complementary
velocity formulation.

\section{Canonical phase space and constraint algebra}
\label{sec:canonical}

The material condition of Sec.~\ref{sec:abelian} selects the
material sector directly, but it does not by itself display which
directions in an enlarged canonical phase space are gauge and which are
removed by second-class constraints.  This distinction is essential:
tangential motion should generate only a relabeling of the interface,
whereas normal motion changes the occupied Hall domain and must remain
physical.  It is also needed to distinguish the material sector from
the larger velocity formulation, which retains independently stored
interfacial charge.  We therefore analyze the velocity completion using
the Dirac--Bergmann procedure~\cite{Dirac,HenneauxTeitelboim,FaddeevJackiw1988}. We use the enlarged phase space
\begin{equation}
(X^i,P_i;\phi,\Pi_\phi;\lambda,\pi_\lambda).
\label{eq:enlarged_phase_space}
\end{equation}
Canonical Poisson brackets are written as $\{\cdot,\cdot\}$, brackets after second-class reduction as $\{\cdot,\cdot\}_D$, and weak equality on the constraint surface as $\approx0$.

\subsection{Auxiliary velocity-constrained completion}

The equal-time condition already selects the material sector.  For canonical classification it is useful to introduce an auxiliary formulation that imposes only the velocity relation and therefore retains excess interfacial charge.  Working in the normal-extension gauge of Eq.~\eqref{eq:normal_extension_gauge}, define
\begin{equation}
\kappa=\frac{m}{4\pi},
\qquad
c=\frac{B}{m},
\qquad
D_\tau\phi=\dot\phi-\beta\phi'.
\label{eq:kappa_c_definition}
\end{equation}
Introduce an auxiliary worldsheet scalar $\lambda(\tau,\sigma)$ and consider
\begin{equation}
\begin{aligned}
S_{\rm min}
=
\int\dd\tau\,\dd\sigma
\Big\{
&-\kappa\phi'D_\tau\phi
\\
&+\lambda
\left(
\partial_\sigma D_\tau\phi
-c\sqrt{\gamma}\,v_n
\right)
-\cE[\bm X]
\Big\}.
\end{aligned}
\label{eq:minimal_constrained_action}
\end{equation}
Here $\cE[\bm X]=\sqrt{\gamma}\,\varepsilon(K,\partial_sK,\ldots)$ is a one-dimensional density satisfying $\int\dd\sigma\,\cE[\bm X]=E[\bm X]$. Under a time-dependent spatial relabeling, $\phi$ and $\lambda$ transform as scalars, $D_\tau\phi$ and $v_n$ transform as scalars, while $\phi'$, $\partial_\sigma D_\tau\phi$, and $\sqrt{\gamma}v_n$ transform as one-dimensional densities. Thus Eq.~\eqref{eq:minimal_constrained_action} is reparametrization invariant. A variation with respect to $\lambda$ gives
\begin{equation}
\partial_\sigma D_\tau\phi
=
c\sqrt{\gamma}\,v_n,
\label{eq:lambda_equation}
\end{equation}
which is Eq.~\eqref{eq:implicit_constraint} in the chosen extension gauge.  For a closed interface in a fixed winding and charge sector, its integral gives $\oint\dd s\,v_n=0$.  The fixed-area condition therefore follows without a separate multiplier, although compact winding and global charge modes still require an independent global specification.

After integrating the $\lambda\partial_\sigma D_\tau\phi$ term by parts along a closed interface,
\begin{equation}
S_{\rm min}
=
\int\dd\tau\,\dd\sigma
\left[
-(\kappa\phi'+\lambda')D_\tau\phi
-c\lambda\sqrt{\gamma}\,v_n
-\cE[\bm X]
\right].
\label{eq:minimal_action_canonical_form}
\end{equation}
This form makes the canonical structure transparent.  The multiplier
$\lambda$ is a reaction field enforcing Hall transport; it is not an
additional edge excitation.  Because this action constrains only the time
derivative of the charge--shape relation, it describes a larger phase space
than the equal-time material reduction.

The same distinction extends to general Abelian interfaces.  The equal-time material constraint selects the sector with no independently stored line charge, whereas the multicomponent velocity completion retains all excess-charge sectors:
\begin{equation}
\begin{aligned}
S_{\Sigma,v}
=
\int\dd\tau\,\dd\sigma\,
\Bigl[&
-\frac{1}{4\pi}
\bm\phi'^TK_\Sigma D_\tau\bm\phi
\\
&+\lambda\left(
\partial_\sigma D_\tau\varphi_c
-B\Delta\nu\sqrt\gamma\,v_n
\right)
\\
&-\cH_\Sigma-\cE
\Bigr].
\end{aligned}
\label{eq:abelian_velocity_completion}
\end{equation}
The scalar action in Eq.~\eqref{eq:minimal_constrained_action} is recovered
for a Laughlin--vacuum interface.  The multicomponent form will be used below
to include neutral fields in the primary constraint matrix.

\subsection{Canonical momenta and primary constraints}

Using the canonical form of the action,
Eq.~\eqref{eq:minimal_action_canonical_form}, the momenta conjugate to
$\phi$, $\bm X$, and $\lambda$ are
\begin{align}
\Pi_\phi
&=
-(\kappa\phi'+\lambda'),
\label{eq:minimal_Pi_phi}
\\
P_i
&=
(\kappa\phi'+\lambda')
\phi'\frac{X'_i}{\gamma}
-
c\lambda\sqrt{\gamma}\,n_i,
\label{eq:minimal_P_i}
\\
\pi_\lambda
&=0.
\label{eq:minimal_pi_lambda}
\end{align}
These equations are not invertible relations that determine all the
velocities in terms of the momenta.  Instead, they define four primary
constraints:
\begin{align}
\chi
&\equiv
\Pi_\phi+\kappa\phi'+\lambda'
\approx0,
\label{eq:minimal_chi}
\\
\cH_\sigma
&\equiv
P_iX'^i+\Pi_\phi\phi'+\pi_\lambda\lambda'
\approx0,
\label{eq:minimal_Hsigma}
\\
\cC_\perp
&\equiv
P_i n^i+c\lambda\sqrt{\gamma}
\approx0,
\label{eq:minimal_Cperp}
\\
\pi_\lambda
&\approx0.
\label{eq:minimal_pilambda_constraint}
\end{align}

The constraints have distinct roles.  The tangential constraint $\cH_\sigma$ generates changes of the coordinate used to label the interface.  The chiral constraint $\chi$ is the first-order momentum relation that reduces an ordinary scalar to one chiral degree of freedom.  The normal relation $\cC_\perp$ does not generate a gauge displacement; it ties the momentum of a physical normal deformation to the Hall reaction field.  Finally, $\pi_\lambda\approx0$ records that $\lambda$ has no independent kinetic term. The term $\pi_\lambda\lambda'$ in
Eq.~\eqref{eq:minimal_Hsigma} vanishes on the constraint surface, but it
must be retained in the generator because it produces the correct
reparametrization of $\lambda$.

Because the kinetic action is homogeneous of first order in the
velocities, its Legendre transform contains no additional kinetic
energy.  The canonical Hamiltonian is therefore the geometric interface
energy,
\begin{equation}
H_{\rm c}=E[\bm X].
\label{eq:minimal_canonical_H}
\end{equation}
The total Hamiltonian is obtained by adding all primary constraints with
independent multipliers:
\begin{equation}
H_{\rm T}
=
E[\bm X]
+
\int\dd\sigma
\left(
u^\sigma\cH_\sigma
+u^\chi\chi
+u^\perp\cC_\perp
+u^\lambda\pi_\lambda
\right).
\label{eq:minimal_total_H}
\end{equation}
The Dirac--Bergmann consistency conditions require every constraint to
be preserved by the evolution generated by $H_{\rm T}$.  These
conditions either determine some of the multipliers, produce secondary
constraints, or reveal gauge freedom.

\subsection{Spatial relabeling as a first-class gauge symmetry}

The nonvanishing equal-time canonical Poisson brackets are
\begin{align}
\{X^i(\sigma),P_j(\sigma')\}
&=
\delta^i{}_j\delta(\sigma-\sigma'),
\\
\{\phi(\sigma),\Pi_\phi(\sigma')\}
&=
\delta(\sigma-\sigma'),
\\
\{\lambda(\sigma),\pi_\lambda(\sigma')\}
&=
\delta(\sigma-\sigma').
\label{eq:minimal_fundamental_PB}
\end{align}
All other elementary canonical brackets vanish.

To test whether $\cH_\sigma$ generates a relabeling, introduce the
smeared functional
\begin{equation}
G[\xi]
=
\int\dd\sigma\,\xi(\sigma)\cH_\sigma(\sigma).
\label{eq:minimal_Gxi}
\end{equation}
The smearing function $\xi$ is the infinitesimal displacement of the
coordinate labels along the curve.  Acting with $G[\xi]$ through the
Poisson bracket gives
\begin{align}
\{X^i,G[\xi]\}
&=\xi X'^i,
&
\{\phi,G[\xi]\}
&=\xi\phi',
\\
\{\lambda,G[\xi]\}
&=\xi\lambda',
&
\{P_i,G[\xi]\}
&=\partial_\sigma(\xi P_i),
\\
\{\Pi_\phi,G[\xi]\}
&=\partial_\sigma(\xi\Pi_\phi),
&
\{\pi_\lambda,G[\xi]\}
&=\partial_\sigma(\xi\pi_\lambda).
\label{eq:minimal_reparam_actions}
\end{align}
Thus $X^i$, $\phi$, and $\lambda$ transform as worldsheet scalars,
whereas their conjugate momenta transform as one-dimensional densities.
This is precisely the cotangent lift of a spatial reparametrization to
phase space.

The generators close under the Poisson bracket:
\begin{equation}
\{G[\xi_1],G[\xi_2]\}
=
G[\xi_1\xi_2'-\xi_2\xi_1'].
\label{eq:minimal_Witt}
\end{equation}
The expression
\begin{equation}
[\xi_1,\xi_2]_{\rm Lie}
=
\xi_1\xi_2'-\xi_2\xi_1'
\end{equation}
is the Lie bracket of one-dimensional vector fields.  Eq.
\eqref{eq:minimal_Witt} therefore realizes the classical Witt algebra,
or the Lie algebra of Diff$^+(S^1)$ for a closed interface, on the
canonical phase space.

The remaining constraints transform covariantly:
\begin{align}
\{\chi,G[\xi]\}
&=
\partial_\sigma(\xi\chi),
\\
\{\cC_\perp,G[\xi]\}
&=
\partial_\sigma(\xi\cC_\perp),
\\
\{\pi_\lambda,G[\xi]\}
&=
\partial_\sigma(\xi\pi_\lambda).
\label{eq:minimal_covariant_constraints}
\end{align}
Their Poisson brackets with $G[\xi]$ are therefore proportional to
constraints and vanish weakly.  Moreover, the geometric energy is
independent of the parametrization of the curve:
\begin{equation}
\{G[\xi],E[\bm X]\}=0.
\label{eq:G_energy_zero}
\end{equation}
It follows that $\cH_\sigma$ is first class. Its arbitrary multiplier $u^\sigma$ reflects a genuine gauge freedom: the equations of motion do not determine how coordinate labels move tangentially along the interface.

\subsection{Charge–shape constraint and the perturbative second-class sector}

The remaining constraints do not generate independent gauge transformations. Their elementary Poisson brackets are
\begin{align}
\{\chi(\sigma),\chi(\sigma')\}
&=
2\kappa\partial_\sigma\delta(\sigma-\sigma'),
\label{eq:minimal_chichi}
\\
\{\chi(\sigma),\pi_\lambda(\sigma')\}
&=
\partial_\sigma\delta(\sigma-\sigma'),
\label{eq:minimal_chipilambda}
\\
\{\cC_\perp(\sigma),\pi_\lambda(\sigma')\}
&=
c\sqrt{\gamma(\sigma)}\delta(\sigma-\sigma'),
\label{eq:minimal_Cpilambda}
\\
\{\chi(\sigma),\cC_\perp(\sigma')\}
&=0.
\label{eq:minimal_chiC}
\end{align}
These brackets are entries of the Dirac matrix. Their nonvanishing values mean that the constraints impose phase-space locking rather than independent gauge redundancies.

The bracket of two smeared normal constraints contains an additional geometric entry that must be retained. Define
\begin{equation}
D_X[\xi]
=
\int\dd\sigma\,\xi P_iX'^{i},
\qquad
\cC_\perp[\eta]
=
\int\dd\sigma\,\eta\cC_\perp.
\end{equation}
A direct canonical calculation gives
\begin{equation}
\{\cC_\perp[\eta],\cC_\perp[\zeta]\}
=
D_X\left[
\frac{\zeta\eta'-\eta\zeta'}{\gamma}
\right].
\label{eq:minimal_CC}
\end{equation}
Geometrically, this bracket realizes the commutator of two normal deformations, whose antisymmetrized action is tangential. The terms proportional to $\lambda\sqrt\gamma$ cancel.

To examine the local nonzero-mode sector, define
\begin{equation}
\Psi_A=(\chi,\cC_\perp,\pi_\lambda),
\quad
C_{AB}(\sigma,\sigma')
=
\{\Psi_A(\sigma),\Psi_B(\sigma')\}.
\label{eq:Dirac_matrix_definition}
\end{equation}
Introduce
\begin{equation}
q(\sigma)
=
\frac{P_iX'^i}{\gamma}.
\label{eq:q_tangential_momentum}
\end{equation}
Eq.~\eqref{eq:minimal_CC} implies that the normal-normal entry acts on a test function as
\begin{equation}
C_{\perp\perp}
=
-\bigl(2q\partial_\sigma+q'\bigr).
\label{eq:Cperpperp_operator}
\end{equation}
Freezing the smooth coefficients and neglecting their gradients, the local Fourier matrix is
\begin{equation}
\mathbb C(k)
=
\begin{pmatrix}
2\ii\kappa k & 0 & \ii k\\
0 & -2\ii qk & c\sqrt\gamma\\
-\ii k & -c\sqrt\gamma & 0
\end{pmatrix}.
\label{eq:corrected_constraint_symbol}
\end{equation}
Although the displayed fixed-$k$ matrix is $3\times3$, there is no conflict
with the vanishing determinant of a real odd-dimensional antisymmetric matrix:
reality pairs the $k$ and $-k$ sectors, and $\mathbb C(k)^T=-\mathbb C(-k)$.
Its determinant is
\begin{equation}
\det\mathbb C(k)
=
2\ii k
\left(
\kappa c^2\gamma+qk^2
\right).
\label{eq:corrected_constraint_determinant}
\end{equation}
The normal-deformation bracket and the resulting local determinant are
derived in Appendix~\ref{app:minimal_constraints}.

For the straight equilibrium configuration, $q=0$ and $\gamma=1$, so
\begin{equation}
\det\mathbb C(k)
=
2\ii\kappa c^2 k
\neq0
\qquad(k\neq0).
\label{eq:equilibrium_constraint_determinant}
\end{equation}
Moreover, $q$ begins beyond linear order about that background. The nonzero Fourier modes therefore form a second-class system in the linear theory, and this invertibility persists perturbatively for sufficiently small amplitude and momentum within the effective-theory regime. Eq.~\eqref{eq:corrected_constraint_determinant} also shows why a global claim on an arbitrary nonlinear background would be too strong: at finite $q$ the operator can, in principle, develop additional kernels.

Where the Dirac matrix is invertible, the reduced bracket is
\begin{widetext}
\begin{equation}
\{F,G\}_D
=
\{F,G\}
-
\int\dd\sigma\,\dd\sigma'\,
\{F,\Psi_A(\sigma)\}
(C^{-1})^{AB}(\sigma,\sigma')
\{\Psi_B(\sigma'),G\}.
\label{eq:general_Dirac_bracket}
\end{equation}
\end{widetext}
After replacing Poisson brackets by Dirac brackets, the second-class constraints may be imposed strongly. The explicit nonlinear inverse is not needed for the perturbative classification; the small-deformation sector gives an explicit reduced bracket below.

The local Fourier analysis does not determine compact-boson winding, total charge, or every zero mode on a circle.  At $k=0$ the derivative entries vanish: the zero mode of $\chi$ generates the constant shift of $\phi$, whereas the zero modes of $\cC_\perp$ and $\pi_\lambda$ remain paired through Eq.~\eqref{eq:minimal_Cpilambda}.  Compact winding and total charge therefore require a separate global analysis.  Locally, $\cH_\sigma$ is first class, while the nonzero modes of $(\chi,\cC_\perp,\pi_\lambda)$ are second class about equilibrium and perturbatively nearby.  There is no independent first-class normal-deformation constraint: normal displacement changes the occupied fluid domain, and $\cC_\perp$ instead locks its momentum to the Hall reaction field.  This differs from the relativistic parent formulation of the Floreanini--Jackiw boson, where both Hamiltonian and spatial worldsheet constraints are first class before gauge fixing~\cite{Townsend2020}.  Here laboratory time is physical, so no first-class time-reparametrization constraint is introduced.

\subsection{Complete multicomponent nonzero-mode constraint matrix}

The same analysis extends directly to the folded multicomponent theory.  For the velocity completion in Eq.~\eqref{eq:abelian_velocity_completion}, let $A,B=1,\ldots,N_\Sigma$ label the folded bosons.  After integrating the $\lambda\partial_\sigma D_\tau\varphi_c$ term by parts, the primary constraints are
\begin{align}
\chi_A
&\equiv
\Pi_A
+\frac{1}{4\pi}K_{\Sigma,AB}\phi'^B
+t_{\Sigma,A}\lambda'
\approx0,
\label{eq:multi_chiral_constraints}
\\
\cH_\sigma
&\equiv
P_iX'^i+\Pi_A\phi'^A+\pi_\lambda\lambda'
\approx0,
\label{eq:multi_reparam_constraint}
\\
\cC_\perp
&\equiv
P_in^i+B\Delta\nu\,\lambda\sqrt\gamma
\approx0,
\label{eq:multi_normal_constraint}
\\
\pi_\lambda&\approx0.
\label{eq:multi_lambda_constraint}
\end{align}
The interface Hamiltonian $\cH_\Sigma$ does not alter these primary momentum relations.  Its role is to propagate or gap the reduced charged and neutral modes. The nonvanishing elementary brackets are
\begin{align}
\{\chi_A(\sigma),\chi_B(\sigma')\}
&=
\frac{1}{2\pi}K_{\Sigma,AB}
\partial_\sigma\delta(\sigma-\sigma'),
\label{eq:multi_chichi}
\\
\{\chi_A(\sigma),\pi_\lambda(\sigma')\}
&=
t_{\Sigma,A}
\partial_\sigma\delta(\sigma-\sigma'),
\label{eq:multi_chipilambda}
\\
\{\cC_\perp(\sigma),\pi_\lambda(\sigma')\}
&=
B\Delta\nu\sqrt\gamma\,
\delta(\sigma-\sigma'),
\label{eq:multi_Cpilambda}
\end{align}
with $\{\chi_A,\cC_\perp\}=0$.  The normal-normal entry remains Eq.~\eqref{eq:Cperpperp_operator}.  Freezing the smooth coefficients, the complete local Fourier matrix in the ordered basis $(\chi_A,\cC_\perp,\pi_\lambda)$ is
\begin{equation}
\mathbb C_\Sigma(k)
=
\begin{pmatrix}
\dfrac{\ii k}{2\pi}K_\Sigma & 0 & \ii k\bm t_\Sigma\\
\\
0 & -2\ii qk & B\Delta\nu\sqrt\gamma\\
\\
-\ii k\bm t_\Sigma^T & -B\Delta\nu\sqrt\gamma & 0
\end{pmatrix}.
\label{eq:multicomponent_constraint_matrix}
\end{equation}
Using the Schur complement and Eq.~\eqref{eq:folded_anomaly_coefficient},
\begin{equation}
\begin{aligned}
\det\mathbb C_\Sigma(k)
=
&\left(\frac{\ii k}{2\pi}\right)^{N_\Sigma}
\det K_\Sigma
\\
&\times
\left[
B^2\Delta\nu^2\gamma
+4\pi q\Delta\nu\,k^2
\right].
\end{aligned}
\label{eq:multicomponent_constraint_determinant}
\end{equation}
The multicomponent Schur-complement calculation is also given in
Appendix~\ref{app:minimal_constraints}.
Fig.~\ref{fig:constraint_reduction} summarizes the two-stage canonical
reduction.  Spatial relabeling is removed as a first-class gauge redundancy,
whereas the chiral, normal, and auxiliary nonzero-mode constraints are
eliminated through the Dirac bracket.

\begin{figure*}[t]
\centering
\includegraphics[width=0.9\textwidth]
{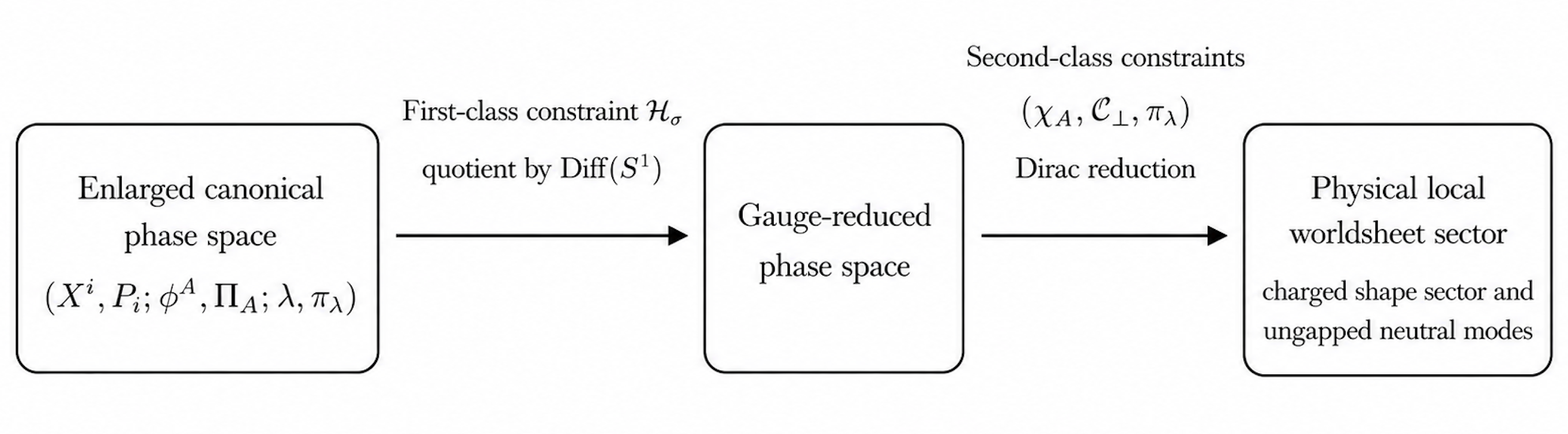}
\caption{
Two-stage local canonical reduction.  The first-class constraint
$\mathcal H_\sigma$ generates spatial relabeling and is removed by
quotienting by $\mathrm{Diff}^+(S^1)$.  About a straight equilibrium
interface with $\Delta\nu\neq0$, the nonzero modes of
$(\chi_A,\mathcal C_\perp,\pi_\lambda)$ form a second-class system and
are eliminated through the Dirac bracket.  After fixing the nonpropagating
conserved profiles retained by the auxiliary completion, the resulting local
worldsheet sector contains the charged shape sector together with any neutral
modes not gapped by allowed interface interactions.
}
\label{fig:constraint_reduction}
\end{figure*}

About the straight equilibrium interface, $q=0$ and $\gamma=1$.  Hence, for $\Delta\nu\neq0$ and $k\neq0$, the full charged, neutral, normal, and auxiliary nonzero-mode matrix is invertible.  The conclusion persists perturbatively in the long-wavelength regime.  Neutral fields are therefore not extra gauge redundancies. After the second-class chiral constraints are eliminated, they remain as physical interface modes whose gaps are controlled by Eq.~\eqref{eq:abelian_gapping_potential}.  For $\Delta\nu=0$, the shape--charge block is singular, in agreement with the absence of a shape Poisson structure generated by the electromagnetic Chern–Simons response.

In the charged--neutral basis of Eq.~\eqref{eq:charged_neutral_decomposition}, the matrix separates into neutral Floreanini--Jackiw blocks and a scalar charged--shape block.  A one-field shape theory is therefore complete only when every neutral mode is either gapped or explicitly retained in the low-energy worldsheet spectrum.

In the material-sector reduction, the velocity multiplier is unnecessary.  One first imposes the chiral constraints $\chi_A$ and obtains the folded current algebra in Eq.~\eqref{eq:abelian_charge_KM}.  The equal-time constraint $\cQ_{\rm mat}=0$ then identifies the charged density with the relative-area density.  In linear static gauge this gives
\begin{equation}
\varphi_c'=B\Delta\nu\,u,
\end{equation}
and the induced bracket is immediately Eq.~\eqref{eq:abelian_shape_bracket}.  Because $\cQ_{\rm mat}$ itself, rather than only its time derivative, is imposed, no arbitrary excess-charge profile appears in this material-sector reduction.

While the constraint analysis above determines the local physical degrees of
freedom, it does not yet express their dynamics in the familiar
small-deformation variables.  We next select the material sector,
perform the linear static-gauge reduction, and verify that the resulting
bracket and spectrum reproduce the established long-wavelength
interface theory.

\section{Linear reduction and interface dynamics}
\label{sec:gaugefix}

The canonical analysis establishes the local degree-of-freedom count.
We now test the construction dynamically.  After selecting the material
sector and expanding about a straight interface, the material and
auxiliary velocity formulations must yield the same reduced bracket and reproduce
the established long-wavelength spectrum.  The reduction also makes
explicit the additional conserved profile retained by the enlarged
velocity theory.

\subsection{Material charge and the linear bracket}

For the static gauge $\bm X=(x,u)$, one has
\begin{equation}
\sqrt{\gamma}\,v_n=\dot u
\label{eq:static_sweep}
\end{equation}
exactly.  At linear order, the Eulerian and material derivatives coincide at the boundary, and Eq.~\eqref{eq:implicit_constraint} becomes
\begin{equation}
\partial_x\partial_t\phi
=
\frac{B}{m}\partial_tu.
\label{eq:linear_constraint_time}
\end{equation}
After fixing the time-independent background profile and removing the conserved area mode,
\begin{equation}
\partial_x\phi
=
\frac{B}{m}u
=
2\pi\rho_0u.
\label{eq:linear_phi_u}
\end{equation}
This is precisely the displacement--density relation used in Ref.~\cite{LiMa2021}.

At this order Eq.~\eqref{eq:implicit_topological_action} becomes the first-order kinetic, or symplectic, term of the Floreanini--Jackiw action~\cite{FloreaniniJackiw}
\begin{equation}
S_{\rm top}^{(2)}
=
-\frac{m}{4\pi}
\int\dd t\,\dd x\,
\phi'\dot\phi.
\label{eq:FJ_linear}
\end{equation}
Its field-space potential and two-form are
\begin{align}
\Theta_\phi
&=
-\frac{m}{4\pi}
\int\dd x\,
\phi'\delta\phi,
\\
\Omega_\phi
&=
-\frac{m}{4\pi}
\int\dd x\,
\partial_x\delta\phi\wedge\delta\phi.
\label{eq:FJ_symplectic_forms}
\end{align}
The constant mode is null.  On the zero-mode-free subspace,
\begin{equation}
\{\phi(x),\phi(y)\}_D
=
\frac{\pi}{m}\operatorname{sgn}_0(x-y),
\label{eq:phi_Dirac_bracket}
\end{equation}
The inversion on the zero-mode-free periodic subspace is given in
Appendix~\ref{app:FJ}.
Therefore,
\begin{equation}
\left\{
\frac{\phi'(x)}{2\pi},
\frac{\phi'(y)}{2\pi}
\right\}_D
=
-\frac{1}{2\pi m}\partial_x\delta(x-y).
\label{eq:KM_bracket}
\end{equation}
Using Eq.~\eqref{eq:linear_phi_u}, the reduced shape bracket is
\begin{equation}
\{u(x),u(y)\}_D
=
-\frac{2\pi m}{B^2}\partial_x\delta(x-y).
\label{eq:u_Dirac_bracket}
\end{equation}
Eqs.~\eqref{eq:FJ_symplectic_forms}--\eqref{eq:u_Dirac_bracket} are exact within the linear small-deformation reduction.  They are not assumed to remain unchanged at finite amplitude.

\subsection{Gauge, boundary-current, and projected-density algebras}

Having obtained the physical boundary-density bracket explicitly, we
can now distinguish it from both the gauge algebra of spatial
relabeling and the projected-density algebra of the bulk Hall fluid.

Eq.~\eqref{eq:minimal_Witt} is the gauge algebra of worldsheet relabelings and acts only on the coordinate labels.  The reduced boundary density obeys the physical $U(1)$ Kac--Moody algebra in Eq.~\eqref{eq:KM_bracket}.  Projected bulk densities instead satisfy the quantum GMP commutator~\cite{GMP1986}
\begin{equation}
[\bar\rho_{\bm q},\bar\rho_{\bm p}]
=
2\ii
\sin\left(
\frac{\ell_B^2}{2}\bm q\times\bm p
\right)
\bar\rho_{\bm q+\bm p}.
\label{eq:GMP}
\end{equation}
Its long-wavelength moments are related to quantum area-preserving
deformations and to $W_\infty$ or $\Walg$ descriptions of QH droplets
and edges~\cite{IsoKarabaliSakita,Cappelli1993,Cappelli1996,CappelliMaffi2021}. These three structures should not be identified:
\begin{itemize}
\item  $\mathrm{Diff}^+(S^1)$ is a gauge algebra,
\item  $U(1)$\text{ Kac--Moody} is the reduced boundary-current algebra,
\item GMP and $W_\infty$ organize physical projected-density
observables in the two-dimensional Hall fluid.
\end{itemize}

In the linearized regime,
\begin{equation}
\delta\rho(x,y)
\simeq
\rho_0u(x)\delta(y)
\label{eq:boundary_density_linear}
\end{equation}
relates the shape bracket to the long-wavelength boundary reduction of
the bulk density algebra.  Nonlinear droplet algebras are known in phase-space descriptions, but
constructing the corresponding local two-sided material-interface map within
the present folded $K$-matrix framework remains an open problem.

\subsection{Velocity theory and excess charge}

The physical meaning of the above classification is most transparent
in the small-deformation regime.  In static gauge and to quadratic
order, the constrained action becomes
\begin{equation}
S_{\rm min}^{(2)}
=
\int\dd t\,\dd x
\left[
-\kappa\phi'\dot\phi
+
\lambda(\dot\phi'-c\dot u)
\right]
-
\int\dd t\,H_2[u].
\label{eq:minimal_linear_action}
\end{equation}
The equations obtained by varying $\lambda$, $\phi$, and $u$ are
\begin{align}
\partial_t(\phi'-cu)
&=0,
\label{eq:minimal_linear_lambda_eq}
\\
\partial_t(2\kappa\phi'+\lambda')
&=0,
\label{eq:minimal_linear_phi_eq}
\\
c\dot\lambda
&=
\frac{\delta H_2}{\delta u}.
\label{eq:minimal_linear_u_eq}
\end{align}
The first equation conserves the difference between boundary charge and swept-area charge.  Selecting the material sector $\cI_{\rm ex}=0$ reproduces Eq.~\eqref{eq:linear_phi_u}.  The remaining two equations then determine the single physical shape field.  Taking an $x$ derivative of Eq.~\eqref{eq:minimal_linear_u_eq} and using Eqs.~\eqref{eq:minimal_linear_phi_eq} and \eqref{eq:linear_phi_u} gives
\begin{equation}
\dot u
=
-\frac{1}{2\kappa c^2}
\partial_x\frac{\delta H_2}{\delta u}
=
-\frac{2\pi m}{B^2}
\partial_x\frac{\delta H_2}{\delta u}.
\label{eq:minimal_linear_Hamilton_eq}
\end{equation}
This is Hamilton's equation generated by the reduced bracket in Eq.~\eqref{eq:u_Dirac_bracket}.  The velocity completion therefore contains one propagating chiral shape mode and no additional dispersing branch in the linear sector.

The linear velocity completion gives
\begin{equation}
\cI_{\rm ex}(x)
=
\phi'(x)-cu(x),
\qquad
\partial_t\cI_{\rm ex}=0
\label{eq:scalar_excess_profile}
\end{equation}
for the Laughlin--vacuum case.  It is $2\pi$ times the excess line charge not tied to swept area.  It survives because the velocity equation fixes only the time derivative of the charge--shape relation.

A second conserved combination appears before the auxiliary field is eliminated,
\begin{equation}
\cI_\lambda
=
2\kappa\phi'+\lambda',
\qquad
\partial_t\cI_\lambda=0.
\label{eq:lambda_response_profile}
\end{equation}
It fixes the reaction field compatible with the chosen shape trajectory.  It
does not generate a second dispersing branch in the reduced material sector.
A complete interpretation of its zero mode belongs to the global analysis and
is not needed for the local nonzero-mode reduction.

The two formulations describe different sectors.  The displayed velocity completion retains arbitrary but frozen $q_{\rm ex}$ profiles because the multiplier equation gives $\partial_t\cI_{\rm ex}=0$ independently of the Hamiltonian.  It does not implement the general worldsheet continuity equation with a nonzero excess tangential current.  Dynamical excess line charge would require an additional interfacial transport field or current variable.  The material reduction instead imposes
$\cI_{\rm ex}=0$ from the outset.  In that sector the boundary charge is exhausted by swept bulk area, no arbitrary local static profile survives, and the reduced dispersing content is the charged shape mode together with any protected neutral fields.

The multiplier theory is therefore an enlarged theory rather than an alternative definition of the material sector.  The equal-time material condition identifies the sector relevant to a homogeneous self-bound interface, while Eq.~\eqref{eq:excess_line_continuity} describes how additional interfacial charge should be restored when required by the microscopic problem.

\subsection{Dispersion, benchmark, and regime of validity}

In the Monge gauge of Eq.~\eqref{eq:static_embedding}, Eq.~\eqref{eq:linear_phi_u} gives
\begin{equation}
\phi
=
\frac{B}{m}\partial_x^{-1}u
\label{eq:phi_inverse_u}
\end{equation}
with $\partial_x^{-1}$ defined on the zero-average subspace.  Substitution into Eq.~\eqref{eq:FJ_linear} yields
\begin{equation}
S_{\rm top}^{(2)}[u]
=
-\frac{B^2}{4\pi m}
\int\dd t\,\dd x\,
u\,\partial_x^{-1}\dot u.
\label{eq:u_topological_action}
\end{equation}
Its inverse symplectic kernel is Eq.~\eqref{eq:u_Dirac_bracket}.

To quadratic order in amplitude and through four spatial derivatives in the energy,
\begin{equation}
H_2
=
\frac{T_0}{2}
\int\dd x\,u_x^2
+
T_2
\int\dd x\,u_{xx}^2.
\label{eq:H2_k5}
\end{equation}
The functional derivative is
\begin{equation}
\frac{\delta H_2}{\delta u}
=
-T_0u_{xx}+2T_2u_{xxxx}.
\label{eq:H2_derivative}
\end{equation}
Hamilton's equation gives
\begin{equation}
\dot u
=
\frac{2\pi mT_0}{B^2}\partial_x^3u
-
\frac{4\pi mT_2}{B^2}\partial_x^5u.
\label{eq:EOM_k5}
\end{equation}
For $u\propto e^{\ii(kx-\omega t)}$ and the orientation used here,
\begin{equation}
\omega(k)
=
\frac{2\pi m}{B^2}
\left(T_0k^3+2T_2k^5\right)
+O(k^7).
\label{eq:dispersion_k5}
\end{equation}
Reversing the orientation reverses the chirality and the overall sign of $\omega(k)$. Using $\rho_0=B/(2\pi m)$,
\begin{equation}
\frac{2\pi mT_0}{B^2}
=
\frac{T_0}{2\pi m\rho_0^2}.
\label{eq:T0_coefficient_match}
\end{equation}
Identifying $T_0=\sigma$ reproduces the cubic coefficient obtained in Ref.~\cite{LiMa2021}.

Eq.~\eqref{eq:EOM_k5} also makes the physical division of labor transparent.  The Hamiltonian variation of the tension term is proportional to $u_{xx}$, but the Hall Poisson operator contributes an additional $\partial_x$, producing the chiral $u_{xxx}$ evolution.  Likewise, the curvature-squared energy produces $u_{xxxx}$, which the same symplectic operator turns into $u_{xxxxx}$.  The odd derivative order is therefore a consequence of first-order Hall kinematics, not an independent assumption about the dispersion.

The general coefficient can be checked against the solvable integer-QH benchmark.  For the attractive $V_1$ model at $\nu=1$, Ref.~\cite{LiMa2021} obtained
\begin{equation}
T_0
=
\left(\frac{2}{\pi}\right)^{3/2}
\frac{|V_1|}{2\ell_B}.
\label{eq:T0_integer}
\end{equation}
Ref.~\cite{TurkerYang2022} found, restoring dimensions,
\begin{equation}
T_2=-\frac14T_0\ell_B^2
\label{eq:T2_integer}
\end{equation}
for the same microscopic model.  Since $B=\ell_B^{-2}$, Eq.~\eqref{eq:dispersion_k5} becomes
\begin{equation}
\omega(k)
=
\sqrt{\frac{8}{\pi}}|V_1|
\left[
(k\ell_B)^3
-\frac12(k\ell_B)^5
\right]
+O[(k\ell_B)^7].
\label{eq:integer_benchmark}
\end{equation}
This reproduces the numerical coefficient and sign of the large-radius result in Ref.~\cite{TurkerYang2022}.  The factor of two multiplying $T_2$ follows from
\begin{equation}
\frac{\delta}{\delta u(x)}
\int\dd y\,u_{yy}^2
=2u_{xxxx}.
\label{eq:factor_two_T2}
\end{equation}

The benchmark also fixes the regime in which the derivative expansion should be read.  The planar result assumes
\begin{equation}
|u_x|\ll1,
\qquad
|k|\ell_{\rm micro}\ll1.
\label{eq:validity_planar}
\end{equation}
For a closed droplet of radius $R$, a local planar approximation additionally requires $|k|R\gg1$.  Finite-radius terms derived in Ref.~\cite{TurkerYang2022} should not be absorbed into the planar coefficients $T_0$ and $T_2$. Note that a negative $T_2$ does not by itself imply an instability.  The derivative expansion is only asymptotic at small $k$; stabilization at larger momentum depends on higher-gradient terms and microscopic physics.

The locality assumption is equally important.  Long-range Coulomb forces lead to a nonlocal functional of the boundary density and generally to nonanalytic momentum dependence rather than the polynomial series in Eq.~\eqref{eq:dispersion_k5}~\cite{Stone1991,Giovanazzi1994}.  The coefficients $T_i$ should therefore be understood as parameters of a screened, short-range, self-bound interface after gapped bulk modes and the finite interface width have been integrated out.

\section{Geometric derivative expansion and microscopic matching}
\label{sec:nonlinear}

The preceding sections determine the charged kinematics, the local
constraint structure, and the linear reduced dynamics. While the linear reduction fixes the universal Poisson operator, it only determines those geometric coefficients that enter the quadratic Hamiltonian. The remaining input to the effective theory is the nonuniversal geometric Hamiltonian. One must also classify the higher-order geometric operators and identify observables that separate their coefficients. In this section, we organize that Hamiltonian as a derivative expansion and identify observables suitable for microscopic matching.

\subsection{Operator basis and Monge expansion}

After subtracting bulk free-energy contributions and working at phase coexistence or in a fixed-area sector, the most general local geometric energy for a planar interface without additional internal fields is
\begin{equation}
E[\bm X]
=
\int\dd s\,
\cE(K,\partial_sK,\partial_s^2K,\ldots).
\label{eq:general_geometric_energy}
\end{equation}
This is the one-dimensional analogue of the derivative expansions used for elastic curves and membranes~\cite{Helfrich1973,LangerSinger1984}.  The similarity concerns the allowed static invariants only.  The dynamics is not that of an elastic filament with inertia or friction, because the QH interface inherits first-order dynamics from its charged edge sector.

The operator reduction below assumes a smooth closed or periodic interface. Counting $K$ as one inverse length, a convenient basis through fourth geometric order is
\begin{equation}
\begin{split}
E[\bm X]
=
\int\dd s\Bigl[
&T_0+T_1K+T_2K^2+T_3K^3
\\
&+T_{4a}K^4+T_{4b}(\partial_sK)^2
+O(\partial_s^5)
\Bigr].
\end{split}
\label{eq:geometric_basis}
\end{equation}
For a closed curve, integrations by parts give
\begin{align}
\oint\dd s\,K\partial_s^2K
&=-\oint\dd s\,(\partial_sK)^2,
\\
\oint\dd s\,K^2\partial_sK
&=0.
\label{eq:geometric_IBP}
\end{align}
Moreover,
\begin{equation}
\oint\dd s\,K=2\pi w,
\label{eq:turning_number}
\end{equation}
where $w$ is the turning number fixed by orientation.  Thus $T_1$ does not affect local dynamics at fixed topology.

Odd powers of $K$ are allowed for a generic oriented interface between inequivalent phases.  They vanish only when a microscopic symmetry exchanges the two sides while reversing the normal.  In the specific integer model of Ref.~\cite{TurkerYang2022}, particle-hole symmetry enforces the absence of odd powers.

Define
\begin{equation}
p=u_x,
\qquad
q=u_{xx},
\qquad
r=u_{xxx}.
\label{eq:pqr}
\end{equation}
The exact expressions are
\begin{align}
\dd s
&=\sqrt{1+p^2}\,\dd x,
\\
K
&=\frac{q}{(1+p^2)^{3/2}},
\\
\partial_sK
&=\frac{r}{(1+p^2)^2}
-\frac{3pq^2}{(1+p^2)^3}.
\label{eq:exact_Monge_invariants}
\end{align}
Expanding Eq.~\eqref{eq:geometric_basis} through fourth order in the amplitude gives
\begin{equation}
E=E_0+H_2+H_3+H_4+O(u^5),
\label{eq:amplitude_expansion}
\end{equation}
with
\begin{align}
\label{eq:H2_full}
&H_2
=
\int\dd x
\left[
\frac{T_0}{2}p^2+T_2q^2+T_{4b}r^2
\right],
\\ \label{eq:H3_full}
&H_3
=
T_3\int\dd x\,q^3,
\end{align}
and
\begin{equation}
\begin{split}
H_4
=
\int\dd x\Bigl[
&-\frac{T_0}{8}p^4
-\frac{5T_2}{2}p^2q^2
+T_{4a}q^4
\\
&+T_{4b}
\left(-\frac72p^2r^2-6pq^2r\right)
\Bigr].
\end{split}
\label{eq:H4_raw}
\end{equation}
For periodic boundary conditions,
\begin{equation}
\int\dd x\,pq^2r
=
-\frac13\int\dd x\,q^4,
\label{eq:pq2r_IBP}
\end{equation}
so an equivalent quartic basis is
\begin{equation}
\begin{aligned}
H_4
=
\int\dd x\Bigl[
&-\frac{T_0}{8}p^4
-\frac{5T_2}{2}p^2q^2
\\
&+(T_{4a}+2T_{4b})q^4
-\frac{7T_{4b}}{2}p^2r^2
\Bigr].
\end{aligned}
\label{eq:H4_reduced}
\end{equation}
The expansion follows directly from the exact Monge-gauge invariants. Appendix~\ref{app:quartic} gives an independent check.

Using the linear bracket Eq.~\eqref{eq:u_Dirac_bracket}, Eq.~\eqref{eq:H2_full} gives
\begin{equation}
\omega(k)
=
\frac{2\pi m}{B^2}
\left(
T_0k^3+2T_2k^5+2T_{4b}k^7
\right)
+O(k^9).
\label{eq:dispersion_k7}
\end{equation}
Only $T_0$, $T_2$, and $T_{4b}$ can be determined from the linear spectrum.  The coefficients $T_3$ and $T_{4a}$ first enter nonlinear observables.

\subsection{Observables, matching, and scheme dependence}

Eqs.~\eqref{eq:H2_full}--\eqref{eq:H4_raw}, with the equivalent quartic form in Eq.~\eqref{eq:H4_reduced} give the expansion of the static geometric energy through fourth order in the deformation amplitude. Although the material condition fixes the charged density relative to a chosen reference at finite deformation, the explicit
shape-only Dirac bracket obtained after eliminating the charged field
has so far been worked out only in the linear regime.  The displayed
Hamiltonian terms are therefore not, by themselves, complete on-shell
interaction vertices at finite amplitude.  Amplitude-dependent
frequencies and scattering amplitudes may receive contributions both
from the geometric Hamiltonian and from nonlinear corrections generated
by the phase-space reduction.

This issue parallels, but is not identical to, nonlinear hydrodynamics of pinned QH edges.  In those systems a density-dependent velocity and dispersive corrections can lead to Benjamin--Ono or Korteweg--de Vries dynamics, shock formation, and solitons~\cite{Bettelheim2006,Wiegmann2012,NardinCarusotto2023}.  A free interface starts from a different fixed point: the leading linear mode is already cubic, and geometric nonlinearities are tied to actual motion of the boundary rather than only to a density wave on a fixed curve.

Using the Fourier convention
\begin{equation}
u(x)
=
\frac{1}{\sqrt L}\sum_k u_k e^{\ii kx},
\label{eq:Fourier_convention}
\end{equation}
the cubic Hamiltonian may be written in Fourier space as
\begin{equation}
H_3
=
-\frac{T_3}{\sqrt{L}}
\sum_{k_1+k_2+k_3=0}
k_1^2k_2^2k_3^2
u_{k_1}u_{k_2}u_{k_3}.
\label{eq:H3_Fourier}
\end{equation}
For a strictly chiral, asymptotically cubic branch, one-to-two decay is kinematically forbidden at sufficiently small positive momenta because
\begin{equation}
(k_1+k_2)^3-k_1^3-k_2^3
=3k_1k_2(k_1+k_2)
\label{eq:cubic_kinematics}
\end{equation}
vanishes only when one momentum vanishes.  This statement can change at finite momentum if higher-order dispersion or additional branches are relevant.

For
\begin{equation}
u(x)=A\cos(kx),
\qquad
Ak\ll1,
\label{eq:sinusoidal_profile}
\end{equation}
the energy density through $A^4$ is
\begin{equation}
\begin{split}
\frac{\Delta E}{L}
=
&\frac{A^2}{4}
\left(T_0k^2+2T_2k^4+2T_{4b}k^6\right)
\\
&+A^4\Biggl[
-\frac{3T_0}{64}k^4
-\frac{5T_2}{16}k^6
\\
&\hspace{1.5cm}
+\left(
\frac{3T_{4a}}{8}-\frac{9T_{4b}}{16}
\right)k^8
\Biggr]
+O(A^6).
\end{split}
\label{eq:sinusoidal_energy}
\end{equation}
The $T_3$ term vanishes for a single cosine and can be isolated by an asymmetric profile or by three Fourier modes satisfying $k_1+k_2+k_3=0$.

For a circular droplet, write $K_R=s_K/R$ with $s_K=\pm1$ fixed by the orientation and normal convention.  Then
\begin{equation}
\begin{aligned}
E_{\rm circle}(R)
=2\pi\Biggl(&T_0R+s_KT_1+\frac{T_2}{R}
+s_K\frac{T_3}{R^2}
\\
&+\frac{T_{4a}}{R^3}+\cdots\Biggr).
\end{aligned}
\label{eq:circle_energy}
\end{equation}
When radii with different particle numbers are compared, bulk chemical-potential and charging contributions must be removed before Eq.~\eqref{eq:circle_energy} is used to infer interface coefficients. If
\begin{equation}
\omega(k)=\alpha_3k^3+\alpha_5k^5+\alpha_7k^7+\cdots,
\label{eq:alpha_expansion}
\end{equation}
then the linear matching relations are
\begin{equation}
T_0=\frac{B^2}{2\pi m}\alpha_3,
\quad
T_2=\frac{B^2}{4\pi m}\alpha_5,
\quad
T_{4b}=\frac{B^2}{4\pi m}\alpha_7.
\label{eq:linear_matching}
\end{equation}
Static imposed-shape energies are the cleanest probes of $T_3$ and $T_{4a}$ because they do not require knowledge of the nonlinear symplectic form.

Beyond leading order, the numerical values of the curvature coefficients depend on how the microscopic interface is converted into a sharp curve.  One may define the embedding by a fixed-density contour, by a Gibbs dividing surface, or by a moment of the boundary density profile.  A local normal redefinition of that dividing curve reshuffles higher-curvature terms in the effective action, much as a field redefinition reshuffles off-shell operators in an ordinary effective field theory.  Consequently, the geometric convention used to extract $\bm X$ must be stated in any microscopic matching calculation.

The leading tension and physical observables are convention independent, but individual higher-order coefficients need not be.  Robust matching targets are therefore spectra, energy differences between specified shapes, finite-amplitude frequency shifts, and properly normalized matrix elements.  Agreement of several such observables using one interface convention is a stronger test than fitting a single dispersion curve.

\section{Conclusion}
\label{sec:discussion}

We have developed a local kinematic and canonical framework for freely moving
Abelian quantum Hall interfaces.  A normal displacement changes the area
occupied by the two incompressible phases and therefore transfers bulk charge.
The two-sided Chern--Simons response gives the exact jump condition
\begin{equation}
J_{1,n}-J_{2,n}=\Delta\rho\,v_n.
\end{equation}
After a material reference curve and interpolation convention are chosen, the
transgressed magnetic-area one-form integrates this velocity relation into the
equal-time condition
\begin{equation}
\varphi_c'=\Delta\nu\,\mathfrak a_\sigma
\end{equation}
for the sector with no independently stored line charge.  The invariant local
quantity is the excess density
$(\varphi_c'-\Delta\nu\mathfrak a_\sigma)/(2\pi)$; the separate local densities
depend on the chosen dividing convention.

The folded $K$-matrix theory fixes the charged $U(1)$ current algebra through
$\Delta\nu$.  About a straight interface, the material condition converts that
algebra into the shape bracket
\begin{equation}
\{u(x),u(y)\}_D
=-\frac{2\pi}{B^2\Delta\nu}\partial_x\delta(x-y),
\end{equation}
while neutral modes remain physical unless they are removed by allowed local
gapping interactions.  A local geometric Hamiltonian then produces the cubic,
quintic, and higher odd-power terms in the interface dispersion.  The static
geometric expansion also separates coefficients accessible from the linear
spectrum from those requiring nonlinear or imposed-shape observables.

The auxiliary velocity-constrained completion displays the local constraint
structure.  Spatial relabeling is first class, whereas the nonzero charged,
normal, and auxiliary constraints are second class about a straight interface
with $\Delta\nu\neq0$ and remain so perturbatively in the long-wavelength
regime.  The displayed completion retains a frozen excess-charge profile; it
is not a general theory of excess tangential transport.

A reference-independent nonlinear first-order shape action has not been
derived here.  The naive covariantization of the chiral kinetic term is not
off-shell invariant under changes of the relative-area interpolation without
additional electromagnetic boundary terms.  We have therefore removed that
claim rather than assume an incomplete action.  Compact winding sectors,
harmonic Chern--Simons modes, regulated determinants, and geometric responses
beyond the electromagnetic Chern--Simons term also require separate treatment.
These limitations do not affect the linear material bracket or the geometric
long-wavelength expansion, but they are essential for a complete nonlinear
quantum theory.

\begin{acknowledgments}
The author is sincerely grateful to Professor Kun Yang for introducing him to quantum Hall interface physics, and for the insightful discussions and encouragement during his stay at the National High Magnetic Field Laboratory. Although the author has since followed a path outside academia, the present work is a continuation of those discussions, revisited many years later, with the hope of providing a more complete account and systematic framework for the subject.
\end{acknowledgments}

\appendix

\section{Geometric identities for a moving planar curve}
\label{app:moving_geometry}

For the velocity decomposition in Eq.~\eqref{eq:velocity_decomposition}, the
line element evolves as
\begin{equation}
\partial_\tau\sqrt{\gamma}
=
\sqrt{\gamma}(\partial_sv_t-Kv_n).
\end{equation}
Consequently,
\begin{equation}
\frac{\dd L}{\dd\tau}
=
-\oint\dd s\,K v_n
\end{equation}
for a smooth closed curve.  The Frenet frame obeys
\begin{align}
\partial_\tau\bm t
&=
(\partial_sv_n+Kv_t)\bm n,
\\
\partial_\tau\bm n
&=
-(\partial_sv_n+Kv_t)\bm t,
\end{align}
and
\begin{equation}
[\partial_\tau,\partial_s]
=
(Kv_n-\partial_sv_t)\partial_s.
\end{equation}
Combining these relations gives
\begin{equation}
\partial_\tau K
=
\partial_s^2v_n+K^2v_n+v_t\partial_sK.
\end{equation}
They reduce in Monge gauge to direct differentiations of
Eqs.~\eqref{eq:static_embedding}--\eqref{eq:static_geometry}.

\section{Variations of the line element, curvature, and geometric energy}
\label{app:geometric_variations}

In this appendix, we derive the geometric variation formulas used in
Sec.~\ref{sec:geometry}.  Consider a smooth deformation of the embedded
curve,
\begin{equation}
\delta\bm X
=
\eta_t\bm t+\eta_n\bm n,
\label{eq:app_general_deformation}
\end{equation}
where $\eta_t$ and $\eta_n$ are respectively the tangential and normal
components of the deformation.  The Frenet equations are
\begin{equation}
\partial_s\bm t=K\bm n,
\qquad
\partial_s\bm n=-K\bm t.
\label{eq:app_frenet}
\end{equation}

\subsection{Variation of the line element and arc-length derivative}

The induced metric and line element are
\begin{equation}
\gamma=\bm X'^2,
\qquad
\dd s=\sqrt{\gamma}\,\dd\sigma.
\label{eq:app_metric}
\end{equation}
Their variation follows from
\begin{equation}
\delta\sqrt{\gamma}
=
\frac{\bm X'\cdot\delta\bm X'}{\sqrt{\gamma}}.
\label{eq:app_delta_sqrt_gamma_1}
\end{equation}
Since $\bm X'=\sqrt{\gamma}\,\bm t$ and
$\delta\bm X'=\sqrt{\gamma}\,\partial_s\delta\bm X$, this becomes
\begin{equation}
\delta\sqrt{\gamma}
=
\sqrt{\gamma}\,
\bm t\cdot\partial_s\delta\bm X.
\label{eq:app_delta_sqrt_gamma_2}
\end{equation}
Using Eqs.~\eqref{eq:app_general_deformation} and
\eqref{eq:app_frenet},
\begin{align}
\partial_s\delta\bm X
&=
\partial_s(\eta_t\bm t+\eta_n\bm n)
\nonumber\\
&=
(\partial_s\eta_t)\bm t
+\eta_tK\bm n
+(\partial_s\eta_n)\bm n
-K\eta_n\bm t
\nonumber\\
&=
(\partial_s\eta_t-K\eta_n)\bm t
+
(\partial_s\eta_n+K\eta_t)\bm n.
\label{eq:app_derivative_deformation}
\end{align}
Therefore,
\begin{equation}
\delta(\dd s)
=
(\partial_s\eta_t-K\eta_n)\dd s.
\label{eq:app_delta_ds}
\end{equation}

Because $\partial_s=\gamma^{-1/2}\partial_\sigma$, the variation of the
arc-length derivative is
\begin{align}
\delta\partial_s
=
\delta(\gamma^{-1/2})\partial_\sigma
=
-\frac{\delta\sqrt{\gamma}}{\sqrt{\gamma}}\,
\partial_s.
\end{align}
Using Eq.~\eqref{eq:app_delta_ds},
\begin{equation}
\delta\partial_s
=
-(\partial_s\eta_t-K\eta_n)\partial_s.
\label{eq:app_delta_partial_s}
\end{equation}

For a closed curve, the variation of its total length is consequently
\begin{align}
\delta L
=
\oint\delta(\dd s)
=
\oint\dd s\,
(\partial_s\eta_t-K\eta_n)
=
-\oint\dd s\,K\eta_n,
\label{eq:app_delta_length}
\end{align}
where the tangential term integrates to zero by periodicity.  Thus
tangential deformations only relabel a closed curve, whereas normal
deformations change its physical length.

\subsection{Variation of the Frenet frame and curvature}

The tangent vector is $\bm t=\partial_s\bm X$.  Its variation must include
the variation of the differential operator:
\begin{align}
\delta\bm t
=
\delta(\partial_s\bm X)
=
(\delta\partial_s)\bm X
+\partial_s\delta\bm X.
\label{eq:app_delta_t_start}
\end{align}
Substituting Eqs.~\eqref{eq:app_derivative_deformation} and
\eqref{eq:app_delta_partial_s}, the tangential contributions cancel and
one obtains
\begin{equation}
\delta\bm t
=
(\partial_s\eta_n+K\eta_t)\bm n.
\label{eq:app_delta_t}
\end{equation}
Preservation of $\bm t\cdot\bm n=0$ and
$\bm n\cdot\bm n=1$ then gives
\begin{equation}
\delta\bm n
=
-(\partial_s\eta_n+K\eta_t)\bm t.
\label{eq:app_delta_n}
\end{equation}

The signed curvature is
\begin{equation}
K=\bm n\cdot\partial_s\bm t.
\label{eq:app_K_definition}
\end{equation}
Its variation is
\begin{align}
\delta K
&=
\delta\bm n\cdot\partial_s\bm t
+
\bm n\cdot\delta(\partial_s\bm t).
\label{eq:app_delta_K_start}
\end{align}
The first term vanishes because
$\delta\bm n$ is proportional to $\bm t$, whereas
$\partial_s\bm t=K\bm n$.  For the second term,
\begin{align}
\delta(\partial_s\bm t)
&=
(\delta\partial_s)\bm t
+
\partial_s\delta\bm t.
\label{eq:app_delta_dst}
\end{align}
Using Eqs.~\eqref{eq:app_delta_partial_s} and
\eqref{eq:app_delta_t},
\begin{align}
\delta K
=
-K(\partial_s\eta_t-K\eta_n)
+\partial_s(\partial_s\eta_n+K\eta_t).
\label{eq:app_delta_K_intermediate}
\end{align}
Expanding the final derivative,
\begin{align}
\delta K
&=
-K\partial_s\eta_t
+K^2\eta_n
+\partial_s^2\eta_n
\nonumber\\
&\quad
+(\partial_sK)\eta_t
+K\partial_s\eta_t.
\end{align}
The two terms proportional to $K\partial_s\eta_t$ cancel, leaving
\begin{equation}
\delta K
=
\partial_s^2\eta_n
+
K^2\eta_n
+
\eta_t\partial_sK.
\label{eq:app_delta_K}
\end{equation}
The last term is the tangential transport of the curvature profile and
does not represent an independent physical deformation of the curve.

\subsection{Variation of the tension and bending energy}

Consider the geometric energy
\begin{equation}
E_{0+2}
=
T_0\oint\dd s
+
T_2\oint\dd s\,K^2.
\label{eq:app_E02}
\end{equation}
Its variation is
\begin{align}
\delta E_{0+2}
={}&
T_0\oint\delta(\dd s)
\nonumber\\
&+
T_2\oint
\left[
K^2\delta(\dd s)
+
2K\,\delta K\,\dd s
\right].
\label{eq:app_delta_E_start}
\end{align}
Substituting Eqs.~\eqref{eq:app_delta_ds} and
\eqref{eq:app_delta_K} gives
\begin{align}
\delta E_{0+2}
=
\oint\dd s\Bigl\{&
T_0(\partial_s\eta_t-K\eta_n)
\nonumber\\
&+
T_2K^2(\partial_s\eta_t-K\eta_n)
\nonumber\\
&+
2T_2K
\left(
\partial_s^2\eta_n
+K^2\eta_n
+\eta_t\partial_sK
\right)
\Bigr\}.
\label{eq:app_delta_E_expanded}
\end{align}

The tangential terms combine into a total derivative:
\begin{align}
&T_0\partial_s\eta_t
+
T_2K^2\partial_s\eta_t
+
2T_2K(\partial_sK)\eta_t
\nonumber\\
=~&
\partial_s
\left[
(T_0+T_2K^2)\eta_t
\right].
\label{eq:app_tangential_total_derivative}
\end{align}
Their integral vanishes for a smooth closed curve.  The remaining normal
variation is
\begin{equation}
\delta E_{0+2}
=
\oint\dd s
\left[
-T_0K\eta_n
+
T_2K^3\eta_n
+
2T_2K\partial_s^2\eta_n
\right].
\label{eq:app_delta_E_normal}
\end{equation}

The last term is integrated by parts twice.  The first integration gives
\begin{equation}
\oint\dd s\,K\partial_s^2\eta_n
=
\left[
K\partial_s\eta_n
\right]_{\partial\Sigma}
-
\oint\dd s\,
(\partial_sK)(\partial_s\eta_n).
\label{eq:app_first_IBP}
\end{equation}
For a closed interface, the boundary term vanishes.  A second integration
by parts gives
\begin{align}
-\oint\dd s\,
(\partial_sK)(\partial_s\eta_n)
&=
-\left[
(\partial_sK)\eta_n
\right]_{\partial\Sigma}
+
\oint\dd s\,
(\partial_s^2K)\eta_n
\nonumber\\
&=
\oint\dd s\,
(\partial_s^2K)\eta_n.
\label{eq:app_second_IBP}
\end{align}
Hence,
\begin{equation}
\oint\dd s\,K\partial_s^2\eta_n
=
\oint\dd s\,
(\partial_s^2K)\eta_n.
\label{eq:app_double_IBP}
\end{equation}
Substituting this result into Eq.~\eqref{eq:app_delta_E_normal}, one obtains
\begin{equation}
\delta E_{0+2}
=
\oint\dd s\,
\left[
-T_0K
+
T_2\left(
2\partial_s^2K+K^3
\right)
\right]\eta_n.
\label{eq:app_delta_E_final}
\end{equation}
This is the normal force generated by the line-tension and
curvature-squared terms.  With the conventions of the main text, the
corresponding functional derivative is
\begin{equation}
\frac{\delta E_{0+2}}{\delta X_n(s)}
=
-T_0K
+
T_2\left(
2\partial_s^2K+K^3
\right).
\label{eq:app_geometric_force}
\end{equation}

\section{Relative-area identities}
\label{app:relative_area}

Let
\begin{equation}
\mathfrak a_a
=
-B\int_0^1\dd r\,
\eps_{ij}Y_r^iY_a^j.
\end{equation}
For constant magnetic two-form $\omega_B(v,w)=B\eps_{ij}v^iw^j$, the closedness of $\omega_B$ gives
\begin{equation}
\partial_\tau\omega_B(Y_r,Y_\sigma)
-
\partial_\sigma\omega_B(Y_r,Y_\tau)
=
\partial_r\omega_B(Y_\tau,Y_\sigma).
\end{equation}
Integrating in $r$, taking the reference curve to be time independent, and using $Y(1)=X$ yields
\begin{equation}
\partial_\tau\mathfrak a_\sigma
-
\partial_\sigma\mathfrak a_\tau
=
-B\eps_{ij}\dot X^iX'^j
=
B\sqrt\gamma\,v_n,
\end{equation}
which proves Eq.~\eqref{eq:relative_area_curvature}.  Stokes' theorem similarly gives Eq.~\eqref{eq:relative_area_integral}.

To compare two homotopies with the same endpoints, let
$\bm Y(\tau,\sigma,r,s)$ interpolate between them, with $0\le s\le1$
and with the endpoint curves fixed as $s$ varies.  Define
\begin{equation}
\Lambda
=
-B\int_0^1\dd s\int_0^1\dd r\,
\eps_{ij}
\partial_rY^i\partial_sY^j.
\end{equation}
Applying the same closed-two-form identity in the $(r,s,a)$ variables gives
\begin{equation}
\mathfrak a_a^{(1)}-\mathfrak a_a^{(0)}
=
\partial_a\Lambda,
\end{equation}
which proves the exact-shift property in
Eq.~\eqref{eq:relative_area_exact_shift}.

\section{Constraint brackets and nonzero-mode determinants}
\label{app:minimal_constraints}

Define
\begin{equation}
N[\eta]
=
\int\dd\sigma\,\eta P_i n^i,
\qquad
A[\eta]
=
c\int\dd\sigma\,\eta\lambda\sqrt\gamma,
\end{equation}
so that $\cC_\perp[\eta]=N[\eta]+A[\eta]$. A normal deformation generated by $N[\eta]$ acts as
\begin{equation}
\delta_\eta X^i=\eta n^i,
\qquad
\delta_\eta n^i=-(\partial_s\eta)t^i.
\end{equation}
Direct functional differentiation gives
\begin{equation}
\{N[\eta],N[\zeta]\}
=
D_X\left[
\frac{\zeta\eta'-\eta\zeta'}{\gamma}
\right].
\end{equation}
The cross terms are symmetric in $\eta$ and $\zeta$,
\begin{equation}
\{N[\eta],A[\zeta]\}
=
c\int\dd\sigma\,
\eta\zeta\lambda\sqrt\gamma\,K,
\end{equation}
and therefore cancel in the antisymmetrized bracket, while $\{A[\eta],A[\zeta]\}=0$. This proves Eq.~\eqref{eq:minimal_CC}.

Writing
\begin{equation}
q=\frac{P_iX'^i}{\gamma},
\end{equation}
Eq.~\eqref{eq:minimal_CC} can be recast as
\begin{equation}
\{\cC_\perp[\eta],\cC_\perp[\zeta]\}
=
\int\dd\sigma\,
q(\zeta\eta'-\eta\zeta').
\end{equation}
After integrating the first term by parts, the associated differential operator is
\begin{equation}
C_{\perp\perp}
=
-(2q\partial_\sigma+q').
\end{equation}
Freezing $q$, $\gamma$, and the remaining smooth coefficients gives the matrix in Eq.~\eqref{eq:corrected_constraint_symbol}. Its determinant is
\begin{equation}
\det\mathbb C(k)
=
2\ii k(\kappa c^2\gamma+qk^2).
\end{equation}
At the straight equilibrium configuration $q=0$ and $\gamma=1$, every nonzero Fourier mode is therefore second class. Since $q$ starts beyond linear order, the same conclusion holds perturbatively at sufficiently small amplitude and momentum. The formula also makes explicit why the argument does not constitute a global invertibility theorem on an arbitrary nonlinear background.

For the multicomponent velocity completion, write the frozen constraint matrix in block form
\begin{equation}
\mathbb C_\Sigma
=
\begin{pmatrix}
A & B_1\\
C_1 & D
\end{pmatrix},
\qquad
A=\frac{\ii k}{2\pi}K_\Sigma.
\end{equation}
Since $K_\Sigma$ is nondegenerate for a valid Abelian Chern--Simons theory, $A$ is invertible for $k\neq0$.  The Schur complement is
\begin{equation}
D-C_1A^{-1}B_1
=
\begin{pmatrix}
-2\ii qk & B\Delta\nu\sqrt\gamma\\
\\
-B\Delta\nu\sqrt\gamma & 2\pi\ii k\Delta\nu
\end{pmatrix}.
\end{equation}
Its determinant is
\begin{equation}
B^2\Delta\nu^2\gamma
+4\pi q\Delta\nu\,k^2,
\end{equation}
which proves Eq.~\eqref{eq:multicomponent_constraint_determinant}.

\section{Inversion of the chiral constraint matrix}
\label{app:FJ}

On the zero-mode-free periodic subspace, define $\operatorname{sgn}_0$ by
\begin{equation}
\partial_x\operatorname{sgn}_0(x-y)
=
2\left[\delta(x-y)-\frac{1}{L}\right].
\label{eq:periodic_sign}
\end{equation}
The inverse of
\begin{equation}
C(x,y)=2\kappa\partial_x\delta(x-y)
\end{equation}
is
\begin{equation}
C^{-1}(x,y)
=
\frac{1}{4\kappa}\operatorname{sgn}_0(x-y).
\end{equation}
The Dirac bracket is therefore
\begin{equation}
\{\phi(x),\phi(y)\}_D
=C^{-1}(x,y)
=\frac{\pi}{m}\operatorname{sgn}_0(x-y),
\end{equation}
which verifies Eq.~\eqref{eq:phi_Dirac_bracket}.

\section{Checks of the quartic geometric expansion}
\label{app:quartic}

The exact static-gauge integrands associated with the operators in
Eq.~\eqref{eq:geometric_basis}, with the common factor $\dd x$ suppressed, are
\begin{align}
\dd s
&\rightarrow
\sqrt{1+p^2},
\\
\dd s\,K^2
&\rightarrow
\frac{q^2}{(1+p^2)^{5/2}},
\\
\dd s\,K^3
&\rightarrow
\frac{q^3}{(1+p^2)^4},
\\
\dd s\,K^4
&\rightarrow
\frac{q^4}{(1+p^2)^{11/2}},
\\
\dd s\,(\partial_sK)^2
&\rightarrow
\frac{r^2}{(1+p^2)^{7/2}}
-\frac{6pq^2r}{(1+p^2)^{9/2}}
+\frac{9p^2q^4}{(1+p^2)^{11/2}}.
\label{eq:exact_operator_densities}
\end{align}
Retaining terms through fourth order in $u$ reproduces the quadratic, cubic, and quartic Hamiltonians displayed in Sec.~\ref{sec:nonlinear}. The final term in Eq.~\eqref{eq:exact_operator_densities} begins at sixth
order and is absent from $H_4$.

\end{document}